\documentclass[runningheads]{llncs}
\usepackage[T1]{fontenc}
\usepackage{graphicx}
\usepackage{amsmath}
\usepackage{amsfonts}
\usepackage{amssymb}
\usepackage{xcolor}
\usepackage{comment}

\usepackage{array}
\usepackage{bbding}

\begin{document}
\def\mi#1{\mathit{#1}}
\title{High-Level Event Mining: \\ Overview and Future Work}
%
\titlerunning{High-Level Event Mining}
%
\author{Bianka Bakullari\,\Envelope\,\inst{1}\orcidID{0000-0003-2680-0826} \and
Wil M. P. van der Aalst \inst{1}\orcidID{0000-0002-0955-6940}}
\authorrunning{B. Bakullari and W.M.P. van der Aalst}
%
\institute{RWTH Aachen University \\
\email{\{bianka.bakullari, wvdaalst\}@pads.rwth-aachen.de}}
%
\maketitle              

\begin{abstract}
Process mining traditionally relies on input consisting of low-level events that capture individual activities, such as filling out a form or processing a product. 
However, many of the complex problems inherent in processes, such as bottlenecks and compliance issues, extend beyond the scope of individual events and process instances. 
Consider congestion, for instance—it can involve and impact numerous cases, much like how a traffic jam affects many cars simultaneously.
High-level event mining seeks to address such phenomena using the regular event data available. 
This report offers an extensive and comprehensive overview at existing work and challenges encountered when lifting the perspective from individual events and cases to system-level events.
\end{abstract}

\section{Motivation}\label{sec:motivation}
Process mining techniques are utilized to analyze and improve processes by using real event data extracted from information systems. 
These data are captured in the form of an event log, which comprises events that occurred during the execution of processes \cite{pm}.
These events involve various tasks, resources, facilities, and costs, among others. 
The attributes of an event specify which of these process components the event concerns. 
In particular, the activity attribute indicates the process task executed during the event. 
Each event is associated with a unique instantiation of the process, as denoted by the case attribute of the event. 
Most process mining methods analyze processes in terms of their process instances, and general claims about the process are often derived from an aggregation of observations at the level of individual cases. 
For example, process performance is assessed by aggregating the durations of individual process instances, and bottlenecks are identified by examining the average time spent between activities. 
However, process behavior is not solely a characteristic of individual cases, as these are not independent from each other. 
Active cases may concurrently demand shared process capacities, potentially overloading the process and workers, leading to costly congestion and delays as the process operates beyond its standard capacity \cite{dirk}. 
Consequently, business processes may encounter various issues due to the fluctuating number of tasks that need to be managed within short time periods. 
This type of behavior can be dynamic in nature; it may emerge locally and be short-lived, thus eluding detection when aggregating entire event data. 

\begin{figure}[t]
\centerline{\includegraphics[scale=0.8]{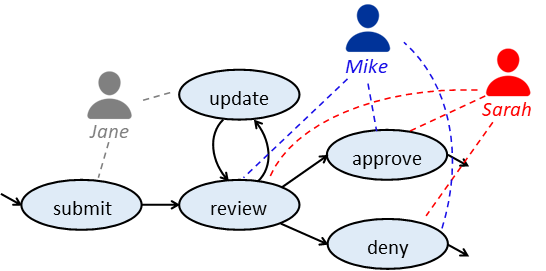}}
\caption{
The citizenship application process involves initial submission, review, optional updates, and decision-making steps at the foreign office. Jane manages submissions and updates, while Mike and Sarah review and decide whether the applications will be approved or denied}
\label{fig: process}
\end{figure}

Consider, for instance, the citizenship application process illustrated in Figure \ref{fig: process}. 
Initially, each applicant submits their citizenship application at the foreign office of their city of residence (activity \emph{submit}). 
Following a waiting period, the application undergoes review (activity \emph{review}). 
If necessary, the applicant may be asked to provide updated or additional documents (activity \emph{update}). 
These are reviewed once more, and finally, a decision is reached, either approving (activity \emph{approve}) or denying (activity \emph{deny}) the citizenship.
In this process, Jane is responsible for managing submission and update-related tasks. 
Mike and Sarah are foreign office workers responsible for reviewing and making decisions on the applications. 
This process exemplifies one where performance can significantly fluctuate based on the frequency of incoming cases.
Suppose, for example, that when applications are assigned for review to Mike in large quantities, they have a higher likelihood of being denied. 
Additionally, the approval process may take longer for applications handled by Sarah if she is very busy when they come up for review. Such patterns may recur during periods of increased application submissions. 
While examining aggregated data over extended periods, fluctuations in waiting times for approval and the proportion of denied cases may not be apparent. 
However, these patterns persist and indicate underlying issues in the process that impact resources, the process runs of active cases, and overall process performance.
Therefore, there is a crucial need to develop new concepts that explicitly capture such behavioral observations to offer a more comprehensive and system-aware perspective of processes. 

In our prior work \cite{hlem}\cite{interplay}, we introduced the concept of "high-level events" to encapsulate and interpret holistic observations, like outlier behavior linked to load and delays. 
This idea of representing such outlier behaviors as events themselves was initially suggested by Zahra et al. \cite{zahra}. 
Additionally, the performance spectrum presented in \cite{perf-spectrum} clearly showed that processes---even within the same segment---exhibit non-stationary behavior which is not observable under aggregation.
This dynamic behavior often appears as batching behavior, as discussed in \cite{eva-batch}, where batching can notably influence performance. Authors in \cite{batch-detection} extended this concept by detecting batching behavior not only within individual tasks but also across linked activities.
Moreover, the authors explored the application of machine learning techniques in \cite{zahra-causal} and \cite{zahra-prescriptive}, utilizing the number of active cases as an additional feature for determining the optimal timing of interventions on running cases. 
In \cite{congestion-graphs}, the authors show how information regarding workload and resource availability can be extracted from raw event data and encoded into congestion graphs. 
From these congestion graphs, congestion-related features can be extracted which are then used for predicting the time until next activity. 
Additionally, the concept of contextual association, introduced in \cite{cross-case}, highlights instances where a group of cases collectively exhibits concept drift in response to a shared object undergoing change. 
Moreover, to address the challenge of analyzing overly fine-grained event data, unsupervised event abstraction techniques have been proposed, as discussed in \cite{chiao-framework} \cite{chiao-po}, allowing for meaningful results by lifting the data to a higher level of abstraction.

This report is structured as follows: 
Section \ref{sec:prelim} lays down fundamental definitions crucial for understanding the proposed methods. 
In Section \ref{sec:hle}, we delve into the concept of high-level events, providing examples and insights into their detection and interpretation. 
Section \ref{sec:cascade and thread} elaborates on connecting individual high-level events through proximity values, leading to the emergence of cascades and threads. 
Section \ref{sec:overlap} demonstrates how to analyze the representation of specific case groups within thread variants. 
In Section \ref{sec:robustness}, we introduce a method to measure process robustness against disruptions by defining the latter as sudden large queues at individual activities. 
Finally, Section \ref{sec:conclusion} concludes the report, touching upon future research directions.

\section{Preliminaries}\label{sec:prelim}
In the remainder, for any set $X$, set $\mathcal{P}(X)$ denotes the power set of $X$, and $X^*$ denotes all sequences over set $X$.
\begin{definition}[Events, Event log]\label{def:event log}
    $\mathcal{U}_{\mi{ev}}$ is the \emph{universe of events} and $\mi{Act}$, $\mi{Case}$, $\mi{Res}$ are the sets of \emph{activity names}, \emph{case identifiers} and \emph{resource names}, respectively.
    $T$ is the totally ordered set of \emph{timestamps}.
    $L=(E,\mi{Attr},\pi)$ is an \emph{event log} where $E\subseteq\mathcal{U}_{\mi{ev}}$ is a finite set of events, ${\{\mi{act}, \mi{case}, \mi{res}, \mi{time}\} \subseteq \mi{Attr}}$ is a set of attribute names and $\pi \in E \times \mi{Attr} \not \to \mi{Val}$ a (partial) function that assigns each event $e$ a value $\pi(e,\mi{att})$ or is undefined (written $\pi(e,\mi{att})\!\!=  \perp$).
    For any $e \in E$, $\pi(e,\mi{act}) \in \mi{Act}$, $ \pi(e,\mi{case}) \in \mi{Case}$, $\pi(e, \mi{res}) \in \mi{Res}$, and $\pi(e,\mi{time}) \in T$.
\end{definition}
I.e., an event log consists of a set of events with various attributes.
The attributes related to the \emph{case}, \emph{activity}, \emph{timestamp}, and \emph{resource} are defined for all events in the log.
For any attribute $\mi{att} \in \mi{Attr}$, for simplicity we write $\mi{att}(e)$ instead of $\pi(e,\mi{att})$.
\begin{definition}[Traces, Steps, Segments]\label{def:steps}
    The cases of an event log $L=(E, \mi{Attr}, \pi)$ are $\mi{cases}(L)=\{\mi{case}(e) \mid e \in E\}$.
    For any case $c \in \mi{cases}(L)$ with corresponding event set $E_c = \{e \in E \mid \mi{case}(e) = c\}$, the \emph{trace} of $c$ is the sequence $\sigma(c)=\langle e_1,...,e_{|E_c|} \rangle \in E_c^*$ containing all events from $E_c$ ordered by time,
    i.e., $\forall_{1 \leq i < j \leq |E_c|} \ \mi{time}(e_i) < \mi{time}(e_j)$.
    A \emph{step} is a pair of directly following events in a case in $L$.
    More precisely, the steps of $L$ are ${\mi{steps}(L)} = \{(e,e') \in E \times E \mid \exists_{c \in \mi{cases}(L)} \ \sigma(c) = \langle ...,e,e',... \rangle \}$.
    Moreover, we define $S(L)=\{(\mi{act}(e),\mi{act}(e')) \mid (e,e') \in \mi{steps}(L)\}$ as the \emph{segments} of $L$.
\end{definition}
A trace is the timely ordered sequence of events that belong to the same case.
A step is a pair of events that occur consecutively within a trace in the log. 
A segment is a pair of activities that directly follow each other in the log.
\begin{definition}[Previous and Next event]\label{def:steps}
    Given event log $L=(E, \mi{Attr}, \pi)$ and the steps set $\mi{steps}(L)$, for any $e \in E$, $\mi{next}(e) = e'$ whenever there is an event $e' \in E$ such that $(e,e') \in \mi{steps}(L)$, and $\mi{next}(e) = \perp$ otherwise.
    Similarly, for any $e \in E$, $\mi{prev}(e) = e'$ whenever there is an event $e' \in E$ such that $(e',e) \in \mi{steps}(L)$, and $\mi{prev}(e) = \perp$ otherwise.
\end{definition}
\begin{definition}[Framing, Time Windows]\label{def:framing}
    A \emph{framing} is a function ${\phi \in T \rightarrow \mathbb{N}}$ mapping timestamps to numbers such that $\forall_{t_1, t_2 \in T} \ t_1 < t_2 \Rightarrow \phi(t_1) \leq \phi(t_2)$.
    Each $w \in \mi{rng}(\phi)$ represents time window  $\overrightarrow{w}=[w_{\mi{start}}, w_{\mi{end}}]$, where $w_{\mi{start}}= \mi{min} \{t \in T \mid \phi(t) = w\}$ and $w_{\mi{end}}= \mi{max} \{t \in T \mid \phi(t) = w \}$.
\end{definition}
Given an event log $L=(E,\mi{Attr},\pi)$ and a framing $\phi$, set $W_{L,\phi}=\{w \in \mathbb{N} \mid 
\mi{min}\{\phi(\mi{time}(e)) \mid e \in E\} \leq w \leq \mi{max}\{\phi(\mi{time}(e)) \mid {e \in E}\} \}$ contains all time windows of $L$ w.r.t. framing $\phi$.
Note that for any $e \in E$, $\phi(\mi{time}(e)) = w$ whenever $e$ occurred within $\overrightarrow{w}$.
In the remainder, we simply refer to $w$ when we mean time window $\overrightarrow{w}$.
For simplicity, we misuse the notation and write $e \in w$ to indicate that $\mi{time}(e) \in \overrightarrow{w}$, we write $e \leq w$ to indicate that $\mi{time}(e) \leq w_{\mi{end}}$, and similarly, $e \geq w$ to indicate that $\mi{time}(e) 
 \geq w_{\mi{start}}$.
 Moreover, we use set $E_w = \{e \in E \mid e \in w\} $ to refer to all events occurring during $w$.
Finally, in the remainder we assume the framing function $\phi$ is fixed and we simply give $W$ as the set of time windows induced by $\phi$ when the log is clear from the context.

Note that the methods described in \cite{hlem}\cite{interplay} can be applied even when the resource attribute is absent from the data, provided that all high-level events requiring resource information are excluded from the analysis. 
Furthermore, all techniques discussed in this report assume the use of a fixed set of time windows, typically of equal size, for observing all high-level events.
\section{High-level Events: Definition and Interpretation} \label{sec:hle}
\subsection{Defining Process Components, Aspects, and High-Level Events}
High-level events are used to conceptualize observations of process behavior at the system level. 
The process aspects typically analyzed concern features like varying process load (congestion), a universal trait across different domains. Congestion aspects can manifest at the levels of activities (e.g., cases waiting in a queue), resources (e.g., resource workload), and segments (e.g., waiting time between two tasks). 
Analogous to system dynamics, various process aspects can be measured in different time windows throughout the process \cite{mahsa}. 
For instance, the number of process instances entering the queue for activity $a$, the number of events handled by resource $r$, and the number of cases entering segment $(a,b)$ are examples of process aspects measurable in any given time window. 
Each time window focuses on a specific set of events related to an individual aspect. 
These events occur close in time (e.g., within the same time window) and in a particular context, which relates to the process aspect, such as referring to the same activity, being handled by the same resource, or indicating the same subsequent activity. 
Depending on the event set linked to a specific aspect in a given time window, there is a real number reflecting the value of that aspect during that window. 
For example, consider an event log $L=(E, \mi{Attr}, \pi)$, where $a$ is an activity and $(a,b)$ is a segment.
The aspect \emph{enqueue} relates to the number of cases entering the queue before a specific activity in a window.
For window $w \in W$ and activity $a$, this event set would be $\{e \in E_w \mid \mi{act}(\mi{next}(e))=a\}$.
A high-level event indicating a large new queue at $a$ during $w$ would emerge when the size of this event set is unusually large, defining the value of the \emph{enqueue} aspect. 
The \emph{handover} aspect focuses on the work handover ratio at a specific segment in a given window.
For window $w \in W$ and segment $(a,b)$, the corresponding event set is $I_{\mi{hd}}=\{e \in E_w \mid \mi{act}(\mi{prev}(e))=a \wedge \mi{act}(e)=b\}$.
A high-level event related to a high work handover ratio at $(a,b)$ during $w$ could arise if the number of resources previously handling activity $a$ is significantly greater than those handling $b$ at $w$, thus determining the value of \emph{handover} at this segment.

Following this, we introduce a unifying definition for all functions that determine the corresponding set of events and the value of various process aspects at the activity, resource, or segment level.
\begin{definition}[Process components]\label{def:comp}
    Given event log $L=(E, \mi{Attr}, \pi)$, let $A(L) = \{\mi{act}(e) \mid e \in E\}$ be the activities of $L$, $R(L) = \{\mi{res}(e) \mid e \in E\}$ be the resources of $L$, and $S(L)$ the segments of $L$. 
    The \emph{components} of $L$ $\mi{comp}(L) = A(L) \cup R(L) \cup S(L)$ refer to the union of all activities, resources and segments in the event log.
\end{definition}
\begin{definition}[Aspect functions]\label{def:aspect}
    $\mathcal{U}_{\mi{asp}}$ is the universe of process aspects.
    Given event log $L=(E, \mi{Attr}, \pi)$ with time windows $W$, for any process aspect $\mi{asp} \in \mathcal{U}_{\mi{asp}}$, there is a corresponding aspect function $f_{\mi{asp}}^L \in \mi{comp}(L) \times W \not \rightarrow \mathcal{P}(E) \times \mathbb{R}$ which (partially) maps component and window pairs of $L$ onto a set of events and a real number.
    In particular, whenever $f_{\mi{asp}}^L \in A(L) \times W  \rightarrow \mathcal{P}(E)\times \mathbb{R}$ we call $\mi{asp}$ an \emph{activity-based aspect}, whenever $f_{\mi{asp}}^L \in R(L) \times W  \rightarrow \mathcal{P}(E)\times \mathbb{R}$ we call $\mi{asp}$ a \emph{resource-based aspect}, and similarly, whenever $f_{\mi{asp}}^L \in S(L) \times W  \rightarrow \mathcal{P}(E)\times \mathbb{R}$ we call $\mi{asp}$ a \emph{segment-based aspect}.
    We drop the superscript $L$ when the event log is clear from the context, and for any component $c \in \mi{comp}(L)$ and window $w \in W$, we write $f_{\mi{asp}}^{\mi{ev}}(c,w)$ and $f_{\mi{asp}}^{\mi{val}}(c,w)$ to refer to the event set and the value of aspect $\mi{asp}$ evaluated at $c$ and $w$, respectively.
\end{definition}
As previously explained, a high-level event related to a specific aspect, component, and time window is detected whenever the value of that aspect, measured for that component at that time window, exceeds a given threshold. 
In the following definition, we limit the types of high-level events to those that pertain to aspects at the activity, resource, and segment levels.
\begin{definition}[High-level event]\label{def:hle}
    Given event log $L=(E, \mi{Attr}, \pi)$ and time window set $W$, let $\mi{ASP} = \mi{ASP_A} \cup \mi{ASP_R} \cup \mi{ASP_S} \subseteq \mathcal{U}_{\mi{asp}}$ be a subset of process aspects where $\mi{ASP_A}$ are activity-based aspects, $\mi{ASP_R}$ are resource-based aspects, and $\mi{ASP_S}$ are segment-based aspects.
    Let $f_{\mi{thresh}} \in (ASP_A \times A) \cup (ASP_R \times R) \cup (ASP_S \times S) \rightarrow \mathbb{R}$ be a threshold mapping which assigns a dedicated threshold to every pair of aspect and component of the corresponding type.
    For any process aspect $\mi{asp} \in \mi{ASP}$, component $c \in \mi{comp}(L)$, and any window $w \in W$, we detect high-level event $h=(\mi{asp},c,w) \in \mi{ASP} \times \mi{comp}(L) \times W$ whenever $f_{\mi{asp}}^{\mi{val}}(c,w) \geq f_{\mi{thresh}}(\mi{asp},c)$.
\end{definition}
\subsection{High-Level Events at the Activity, Resource, and Segment Level}
In the following, we (re)introduce some high-level events referring to various aspects at the activity, resource, and segment levels.
The labels used to describe these aspects in this report are slightly adjusted from those in existing work.
Given an event log $L = (E, \mi{Attr}, \pi)$ and a set of time windows $W$, let $A$, $R$, $S$ represent its sets of activities, resources, and segments, respectively.

Examples of activity-based aspects include $\mi{exec}, \mi{enqueue}, \mi{queue} \in \mathcal{U}_{\mi{asp}}$.
For any activity $a \in A$ and window $w \in W$:
\begin{itemize}
    \item $f_{\mi{exec}}^{\mi{ev}}(a,w) = \{e \in E_w \mid \mi{act}(e) = a\}$ is the set of events executing $a$ during $w$, and $f_{\mi{exec}}^{\mi{val}}(a,w) = |f_{\mi{exec}}^{\mi{ev}}(a,w)|$ is the size of that set.
    A high-level event $(\mi{exec},a,w)$ would indicate that activity $a$ was executed unusually often during $w$.
    \item $f_{\mi{enqueue}}^{\mi{ev}}(a,w) = \{e \in E_w \mid \mi{act}(\mi{next}(e)) = a\} $ is the set of events occurring during $w$ whose next task will be activity $a$. 
    The occurrence of each of these events indicates that the corresponding case enters the queue at $a$ during $w$.
    $f_{\mi{enqueue}}^{\mi{val}}(a,w) = |f_{\mi{enqueue}}^{\mi{ev}}(a,w)|$ is the size of that set.
    A high-level event $(\mi{enqueue},a,w)$ would indicate that a high volume of cases entered the queue for activity $a$ during $w$.
    \item $f_{\mi{queue}}^{\mi{ev}}(a,w) = \{e \in E \mid e \leq w \wedge \mi{next}(e) \geq w  \wedge \mi{act}(\mi{next}(e)) = a\} $ is the set of events that occurred before or during $w$ whose next task will be activity $a$. 
    The occurrence of each of these events indicates that the corresponding case enters the queue at $a$ at some point before or during $w$ and is still in the queue during $w$.
    $f_{\mi{queue}}^{\mi{val}}(a,w) = |f_{\mi{queue}}^{\mi{ev}}(a,w)|$ is the size of that set.
    A high-level event $(\mi{queue},a,w)$ would indicate that a high volume of cases is in the queue for activity $a$ during $w$.
\end{itemize}
In the illustrations showing the event sets of different aspects, we reference the process depicted in Figure \ref{fig: process}. 
Each dot represents an event executing the corresponding activity (determined by the line on the y-axis) at a specific time (positioned on the x-axis). 
The vertical dotted lines split the time scope into time windows. 
A line connecting two events of different activities indicates that the later event directly followed the previous event in the same case.
The darker dots comprise the event set of the particular aspect at a fixed time window $w$.

\begin{figure}[h]
\minipage{0.45\textwidth}
  \includegraphics[width=\linewidth]{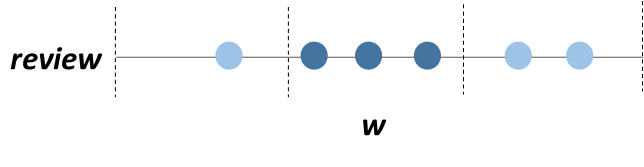}
\endminipage\hfill
\minipage{0.5\textwidth}%
  \includegraphics[width=\linewidth]{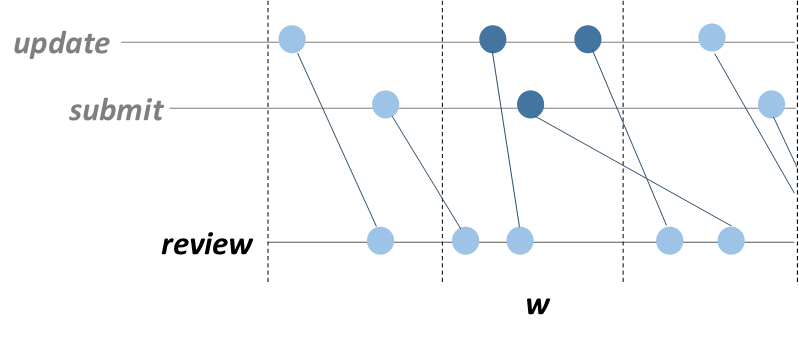}
\endminipage
 \caption{The dark dots (left) correspond to events from set $f_{\mi{exec}}^{\mi{ev}}(\mi{review},w)$ which indicate executions of activity \emph{review} during window $w$.
 The dark dots (right) correspond to events from set $f_{\mi{enqueue}}^{\mi{ev}}(\mi{review},w))$ which indicate cases enqueuing for activity \emph{review} during window $w$.}
 \label{fig: execenqueue}
\end{figure}

\begin{figure}[h]
\centerline{\includegraphics[scale=0.5]{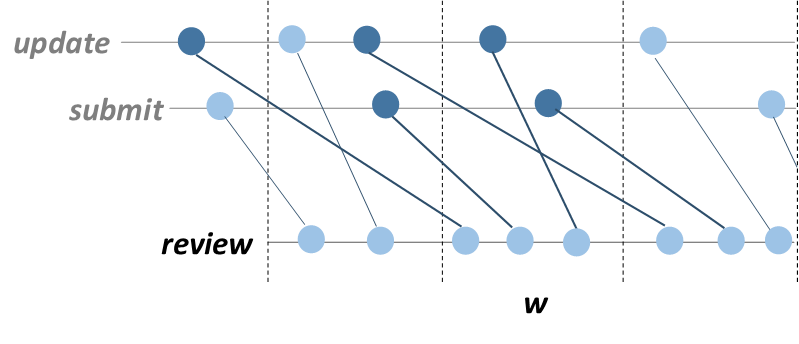}}
\caption{The dark dots correspond to events from set $f_{\mi{queue}}^{\mi{ev}}(\mi{review},w)$ which indicate cases in the queue for activity \emph{review} that have entered the queue before or during window $w$.}
\label{fig: queue}
\end{figure}

In Figure \ref{fig: execenqueue}, the left illustration shows which review events occur during $w$ (event set $f_{\mi{exec}}^{\mi{ev}}(\mi{review},w)$), whereas the right illustration shows which events indicate cases enqueuing for review during $w$ (event set $f_{\mi{enqueue}}^{\mi{ev}}(\mi{review},w))$.
These events can be related to activities \emph{submit} and \emph{update} since they precede \emph{review} in the process.
Figure \ref{fig: queue} depicts the events which indicate the enqueuing at activity \emph{review} before or during $w$ (event set $f_{\mi{queue}}^{\mi{ev}}(\mi{review},w)$).

\begin{figure}[h]
\minipage{0.45\textwidth}
  \includegraphics[width=\linewidth]{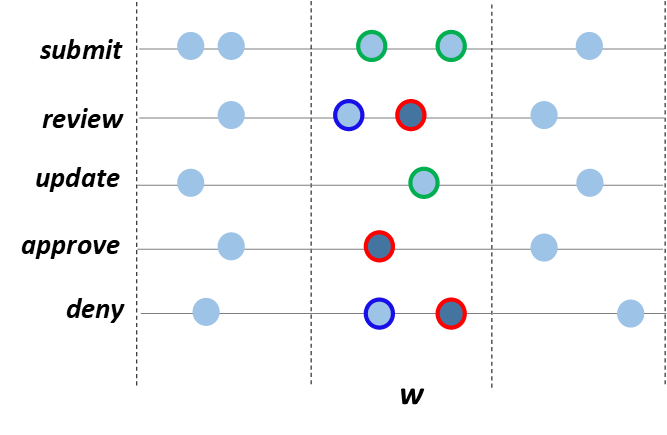}
\endminipage\hfill
\minipage{0.45\textwidth}%
  \includegraphics[width=\linewidth]{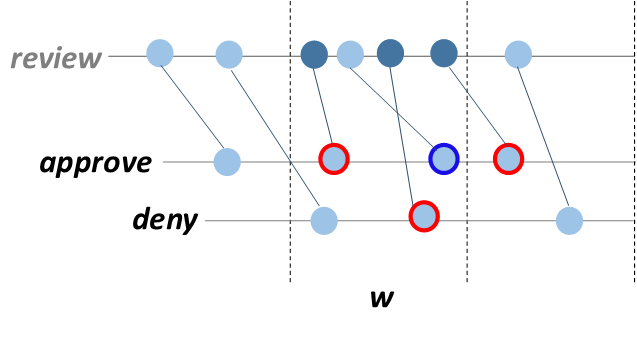}
\endminipage
 \caption{The dots encircled in red in the left illustration represent events that were executed by Sarah during window $w$ (set $f_{\mi{do}}^{\mi{ev}}(\mi{Sarah},w)$).
 The dark dots in the right illustration correspond to a subset of the events occurring during $w$ (specifically \emph{review} events) whose next task (activity \emph{approve} or \emph{deny}) will be handled by Sarah (set $f_{\mi{todo}}^{\mi{ev}}(\mi{Sarah},w)$).}
 \label{fig: dotodo}
\end{figure}

\begin{figure}[h]
\centerline{\includegraphics[scale=0.55]{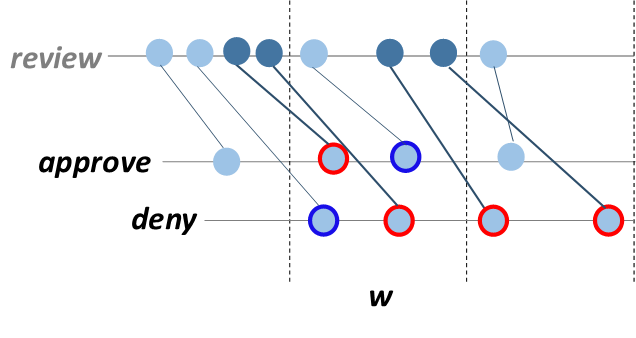}}
\caption{The dots encircled in red represent events that were executed by Sarah. The dark dots correspond to a subset of the events occurring before or during $w$ (specifically \emph{review} events) whose next task (activity \emph{approve} or \emph{deny}) will be handled by Sarah (set $f_{\mi{workload}}^{\mi{ev}}(\mi{Sarah},w)$).}
\label{fig: workload}
\end{figure}

Examples of resource-based aspects include $\mi{do}, \mi{todo}, \mi{workload} \in \mathcal{U}_{\mi{asp}}$.
For any resource $r \in R$ and window $w \in W$:
\begin{itemize}
    \item $f_{\mi{do}}^{\mi{ev}}(r,w) = \{e \in E_w \mid \mi{res}(e) = r\} $ is the set of events handled by resource $r$ during window $w$, and $f_{\mi{do}}^{\mi{val}}(r,w) = |f_{\mi{do}}^{\mi{ev}}(r,w)|$ is the size of that set.
    A high-level event $(\mi{do},r,w)$ would indicate that resource $r$ was particularly busy handling events during $w$.
    \item $f_{\mi{todo}}^{\mi{ev}}(r,w) = \{e \in E_w \mid \mi{res}(\mi{next}(e)) = r\} $ is the set of events occurring during $w$ whose next upcoming task will be handled by resource $r$. 
    The occurrence of each of these events indicates a new task lining up for resource $r$.
    $f_{\mi{todo}}^{\mi{val}}(r,w) = |f_{\mi{todo}}^{\mi{ev}}(r,w)|$ is the size of that set.
    A high-level event $(\mi{todo},r,w)$ would indicate that resource $r$ received a high volume of new tasks during $w$.
    \item $f_{\mi{workload}}^{\mi{ev}}(a,w) = \{e \in E \mid e \leq w \wedge \mi{next}(e) \geq w  \wedge \mi{res}(\mi{next}(e)) = r\} $ is the set of events having occurred before or during $w$ whose next upcoming task will be handled by resource $r$. 
    The occurrence of each of these events indicates that a new task is added to the total workload of $r$ at some point before or during $w$ and is still part of the workload during $w$.
    $f_{\mi{workload}}^{\mi{val}}(r,w) = |f_{\mi{workload}}^{\mi{ev}}(r,w)|$ is the size of that set.
    A high-level event $(\mi{workload},r,w)$ would indicate that there was a high volume of tasks waiting for resource $r$ during $w$.
\end{itemize}
In the illustrations depicting the event sets of different resource-based aspects, the colored circles surrounding the dots representing various events indicate the corresponding resource: gray for Jane, blue for Mike, and red for Sarah. 
Figure \ref{fig: dotodo} (left) shows all events occurring within time window $w$ along with their associated resources. 
The dark dots represent events executed by Sarah during window $w$ (event set $f_{\mi{do}}^{\mi{ev}}(\mi{Sarah},w)$). 
The right illustration displays only a subset of the events that highlight tasks accumulating for Sarah during $w$ (a subset of $f_{\mi{todo}}^{\mi{ev}}(\mi{Sarah},w)$). 
Some of these future tasks may be related to the activities \emph{approve} and \emph{deny}, which typically follow the completion of the prior activity \emph{review}.
Figure \ref{fig: workload} depicts a subset of the events that indicate tasks accumulating for Sarah before or during $w$ (a subset of $f_{\mi{workload}}^{\mi{ev}}(\mi{Sarah},w)$). 
These events represent a portion of Sarah's total workload during $w$.

\begin{figure}[h]
\minipage{0.5\textwidth}
  \includegraphics[width=\linewidth]{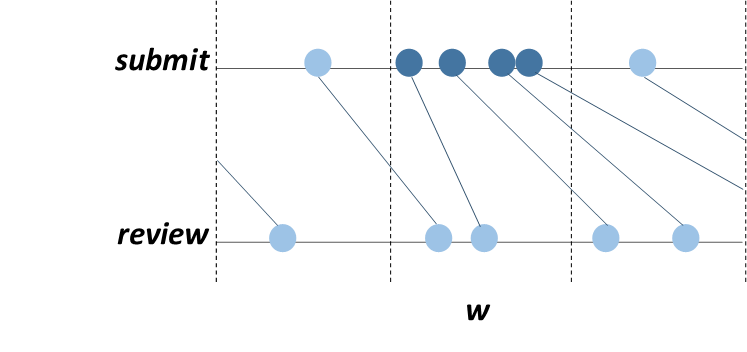}
\endminipage\hfill
\minipage{0.5\textwidth}%
  \includegraphics[width=\linewidth]{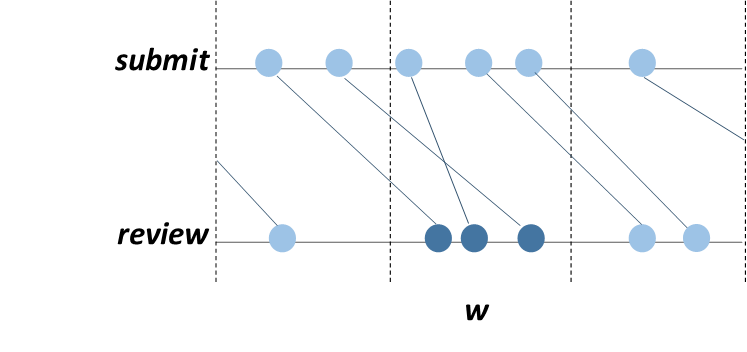}
\endminipage
 \caption{The dark dots (left) correspond to events from set $f_{\mi{enter}}^{\mi{ev}}((\mi{submit},\mi{review}),w)$ which indicate cases entering segment $(\mi{submit},\mi{review})$ during window $w$. The dark dots (right) correspond to events from set $f_{\mi{exit}}^{\mi{ev}}((\mi{submit},\mi{review}),w)$ which indicate cases exiting segment $(\mi{submit},\mi{review})$ during window $w$.}
 \label{fig: enterexit}
\end{figure}

\begin{figure}[h]
\centerline{\includegraphics[scale=0.5]{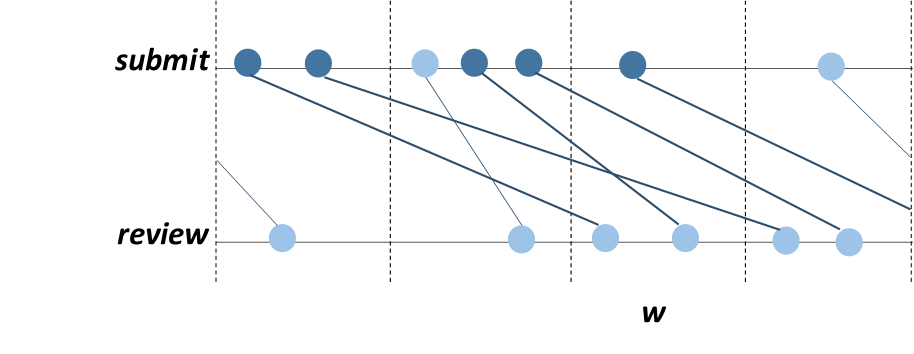}}
\caption{The dark dots correspond to events from set $f_{\mi{cross}}^{\mi{ev}}((\mi{submit},\mi{review}),w)$ which indicate cases crossing segment $(\mi{submit},\mi{review})$ during window $w$.}
\label{fig: cross}
\end{figure}

Examples of segment-based aspects include $\mi{enter}$, $\mi{exit}$, $\mi{cross}$, $\mi{handover}$, \linebreak $\mi{delayStart}$, $\mi{delayEnd}$, $\mi{delayIn}$ $\mi{delayNow} \in \mathcal{U}_{\mi{asp}}$.
For any segment $s=(a,b) \in S$ and window $w \in W$:
\begin{itemize}
    \item $f_{\mi{enter}}^{\mi{ev}}(s,w) = \{e \in E_w \mid \mi{act}(e) = a \wedge \mi{act}(\mi{next}(e))=b\}$ is the set of events executing $a$ during $w$ whose next activity will be $b$. 
    The occurrence of each of these events represents a case entering segment $(a,b)$ during $w$.
    $f_{\mi{enter}}^{\mi{val}}(s,w) = |f_{\mi{enter}}^{\mi{ev}}(s,w)|$ is the size of that set.
    A high-level event $(\mi{enter},s,w)$ would indicate a high volume of cases entering segment $(a,b)$ during $w$.
    \item $f_{\mi{exit}}^{\mi{ev}}(s,w) = \{e \in E_w \mid \mi{act}(\mi{prev}(e)) = a \wedge \mi{act}(e)=b\} $ is the set of events executing $b$ during $w$ whose previous activity was $a$. 
    The occurrence of each of these events represents a case exiting segment $(a,b)$ during $w$.
    $f_{\mi{exit}}^{\mi{val}}(s,w) = |f_{\mi{exit}}^{\mi{ev}}(s,w)|$ is the size of that set.
    A high-level event $(\mi{enter},s,w)$ would indicate a high volume of cases exiting segment $(a,b)$ during $w$.
    \item $f_{\mi{cross}}^{\mi{ev}}(s,w) = \{e \in E \mid e \leq w \wedge \mi{next}(e) \geq w \wedge \mi{act}(e)=a \wedge \mi{act}(\mi{next}(e)) = b\} $ is the set of events having executed $a$ before or during $w$ and whose next task will be $b$. 
    The occurrence of each of these events represents a case crossing segment $(a,b)$ during $w$.
    $f_{\mi{cross}}^{\mi{val}}(s,w) = |f_{\mi{cross}}^{\mi{ev}}(s,w)|$ is the size of that set.
    A high-level event $(\mi{cross},s,w)$ would indicate a high volume of cases crossing segment $(a,b)$ during $w$.
    \item $f_{\mi{handover}}^{\mi{ev}}(s,w) = \{e \in E_w \mid \mi{act}(\mi{prev}(e)) = a \wedge \mi{act}(e)=b\} $ is the set of events executing $b$ during $w$ whose previous activity was $a$. 
    The occurrence of each of these events represents a case exiting segment $(a,b)$ during $w$.
    $f_{\mi{handover}}^{\mi{val}}(s,w) = |\{\mi{res}(\mi{prev}(e)) \mid e \in f_{\mi{handover}}^{\mi{ev}}(s,w)\}| / |\{\mi{res}(e) \mid e \in f_{\mi{handover}}^{\mi{ev}}(s,w)\}|$ reflects the handover ratio at $(a,b)$ during $w$, calculated as the ratio between the number of resources handling the previous activity $a$ and those handling the current activity $b$.
    A high-level event $(\mi{handover},s,w)$ would indicate a significant discrepancy in the number of resources handling these two activities.
    \item $f_{\mi{delayStart}}^{\mi{ev}}(s,w) = \{e \in E_w \mid \mi{act}(e) = a \wedge \mi{act}(\mi{next}(e))=b\}$ is the set of events executing $a$ during $w$ whose next activity will be $b$. 
    Each event represents a case entering segment $(a,b)$ during $w$.
    $f_{\mi{delayStart}}^{\mi{val}}(s,w) = \linebreak \frac{1}{|\mi{f_{\mi{delayStart}}^{\mi{ev}}(s,w)}|} \sum_{e \in f_{\mi{delayStart}}^{\mi{ev}}(s,w)} \big(\mi{time}(\mi{next}(e)) - \mi{time}(e) \big)$ 
    is calculated as the average waiting time between $a$ and $b$ for all cases which entered segment $(a,b)$ during $w$.
    A high-level event $(\mi{delayStart},s,w)$ would indicate that the average waiting time between $a$ and $b$ for cases entering segment $(a,b)$ during $w$ was unusually long.
    \item $f_{\mi{delayEnd}}^{\mi{ev}}(s,w) = \{e \in E_w \mid \mi{act}(\mi{prev}(e)) = a \wedge \mi{act}(e)=b\} $ is the set of events executing $b$ during $w$ whose previous activity was $a$. 
    Each event represents a case exiting segment $(a,b)$ during $w$.
    $f_{\mi{delayEnd}}^{\mi{val}}(s,w) = \linebreak \frac{1}{|\mi{f_{\mi{delayEnd}}^{\mi{ev}}(s,w)}|} \sum_{e \in f_{\mi{delayEnd}}^{\mi{ev}}(s,w)} \big(\mi{time}(e) - \mi{time}(\mi{prev}(e))\big)$ 
    reflects the average waiting time between $a$ and $b$ for all cases exiting segment $(a,b)$ during $w$.
    A high-level event $(\mi{delayEnd},s,w)$ would indicate that the average waiting time between $a$ and $b$ for cases exiting segment $(a,b)$ during $w$ was unusually long.
    \item $f_{\mi{delayIn}}^{\mi{ev}}(s,w) = \{e \in E \mid e \leq w \wedge \mi{next}(e) \geq w \wedge \mi{act}(e)=a \wedge \mi{act}(\mi{next}(e)) = b\} $ is the set of events having executed $a$ before or during $w$ and whose next task will be $b$. 
    Each event represents a case that is crossing segment $(a,b)$ during $w$.
    $f_{\mi{delayIn}}^{\mi{val}}(s,w) = \frac{1}{|\mi{f_{\mi{delayIn}}^{\mi{ev}}(s,w)}|} \sum_{e \in f_{\mi{delayIn}}^{\mi{ev}}(s,w)} \big(\mi{time}(\mi{next}(e)) - \mi{time}(e) \big)$ 
    reflects the average waiting time between $a$ and $b$ for all cases crossing segment $(a,b)$ during $w$.
    A high-level event $(\mi{delayIn},s,w)$ would indicate that the average waiting time between $a$ and $b$ for cases crossing segment $(a,b)$ during $w$ was particularly long.
    \item $f_{\mi{delayNow}}^{\mi{ev}}(s,w) = \{e \in E \mid e \leq w \wedge \mi{next}(e) \geq w \wedge \mi{act}(e)=a \wedge \mi{act}(\mi{next}(e)) = b\} $ is the set of events having executed $a$ before or during $w$ and whose next task will be $b$. 
    Each event represents a case crossing segment $(a,b)$ during $w$.
    The value 
    $f_{\mi{delayNow}}^{\mi{val}}(s,w) = \frac{1}{|\mi{f_{\mi{delayIn}}^{\mi{ev}}(s,w)}|} 
    \bigg(
    \sum_{\substack{e \in f_{\mi{delayIn}}^{\mi{ev}}(s,w) \\ s.t. \ \mi{next}(e) \in w}} \big(\mi{time}(\mi{next}(e)) - \mi{time}(e) \big)
    +
    \sum_{\substack{e \in f_{\mi{delayIn}}^{\mi{ev}}(s,w) \\ s.t. \ \mi{next}(e) > w}} \big(w_{\mi{end}} - \mi{time}(e) \big)
    \bigg)$ 
    is calculated by combining the waiting times between $a$ and $b$ when $b$ occurs within $w$ and adding the waiting times between $a$ and the end of window $w$ when $b$ occurs after $w$.
    A high-level event $(\mi{delayNow},s,w)$ would indicate that the current average waiting time between $a$ and $b$ for the cases crossing segment $(a,b)$ during $w$ was particularly long.
\end{itemize}
We will reference the segment $(\mi{submit},\mi{review})$ from the citizenship application process to illustrate aspects \emph{enter}, \emph{exit}, and \emph{cross}.
In Figure \ref{fig: enterexit}, the left illustration shows the \emph{submit} events of cases whose next activity is \emph{review}. 
These are the events from set $f_{\mi{enter}}^{\mi{ev}}((\mi{submit},\mi{review}),w)$.
In contrast, the right illustration shows the \emph{review} events of cases whose previous activity was \emph{submit}. 
These are the events from set $f_{\mi{exit}}^{\mi{ev}}((\mi{submit},\mi{review}),w)$.
Figure \ref{fig: cross} depicts events of cases which are crossing segment $(\mi{submit},\mi{review})$ during $w$ (event set $f_{\mi{cross}}^{\mi{ev}}((\mi{submit},\mi{review}),w)$).
Specifically, these are the \emph{submit} events that occurred before or during $w$ whose next activity will be \emph{review}.
\subsection{Addressing Challenges in Detecting and Interpreting High-Level Events}
It is important to note that there are four distinct aspects related to the time cases spend waiting in a particular segment. A high-level event related to \emph{delayStart} may occur when cases enter a segment during a particularly busy period, suffering a prolonged wait, although they may not necessarily exit the segment within the same time window. 
Conversely, a high-level event related to \emph{delayEnd} may emerge when multiple cases that have experienced long waits at a segment exit that segment closely together, potentially easing congestion at that time. 
This exit could occur even though they possibly entered the segment at different times.
Furthermore, the aspect \emph{delayIn} measures the average waiting time for all cases crossing a segment within a specific time window, while \emph{delayNow} provides a more operational perspective by showing the average waiting time up until the current time window. 
Depending on the specific questions being addressed, different perspectives on delays may be relevant.
To prevent the identification of high-level events concerning delays that occur simply because a negligible number of cases have extremely long waiting times, it is advisable to include additional criteria that also consider the number of cases involved before declaring a high-level event. 
This approach ensures a more robust and meaningful analysis of delays within process segments.

It is essential to recognize that the aspects mentioned can be categorized into two distinct groups; the aspects that describe an \emph{action} within a given window, and those that describe a \emph{state} at that window.
The first group includes the aspect set $\{\mi{exec}, \mi{enqueue}, \mi{do}, \mi{todo}, \mi{enter}, \mi{exit}, \mi{handover}, \mi{delayStart}, \mi{delayEnd}\}$.
The event set for these aspects always pertains to events that occur specifically within the time window where the aspect function is evaluated.
Consequently, if two high-level events ${h_1=(\mi{asp},c,w_1)}$ and ${h_2=(\mi{asp},c,w_2)}$ refer to the same aspect and component but occur in different time windows, the events responsible for their emergence are always disjoint: $f_{\mi{asp}}^{ev}(c,w_1) \cap f_{\mi{asp}}^{ev}(c,w_2)=\emptyset$. 
In contrast, the second group includes aspect set $\{\mi{queue}, \mi{workload}, \mi{delayIn}, \mi{delayNow}\}$.
These capture the state of the process at a certain point in time, as the events leading to the emergence of the corresponding high-level event do not necessarily belong to the window in which the high-level event is observed.
Thus, when high-level events concerning the same aspect and component occur in subsequent time windows, it may not indicate entirely new behavior, but rather an evolving state of the process as new events are added to and removed from the responsible event set at each window.

In Definition \ref{def:aspect}, each aspect function yields both the value and the underlying event set responsible for computing the value.
It is important to note that for the detection of high-level events, only the value is relevant.
Unless there is a need to access the attributes of the underlying low-level events for the method being used, an explicit computation of the event set is not necessary.
For instance, in \cite{hlem} only the values are relevant for the generation of high-level events.
In \cite{interplay}, the event sets are also computed because the case attribute of these events is relevant later in the method.
Specifically, in \cite{interplay}, aspect functions are referred to as \emph{patterns}, the aspects are called \emph{pattern types}, and the combinations of segment and window pairs are referred to as \emph{coordinates}.

Many aspects take into account the next and previous events of those occurring within the current time window. 
It should be noted that the first events of any case do not have a previous event, and similarly, the last events of any case do not have a next event. 
This leads to situations where, for example, the queues of activities for start events cannot be captured, as the presence of the corresponding case in the log is only recognized after the occurrence of the start event. 

When analyzing high-level behavior, a key challenge is identifying the types of high-level events that are interesting and relevant. 
This involves selecting the appropriate aspects and components, as well as establishing thresholds for detecting high-level events. 
One should ensure that resource-based aspects are only analyzed when the resource attribute is available in the data. 
Furthermore, focusing on aspects that describe \emph{action} rather than \emph{states} can make the interpretation of their recurrence more intuitive.
If the selection of aspects and/or components is not predetermined, or if there is a desire to limit the number of high-level events while retaining the most important ones, it is possible to choose high-level events based on their \emph{coverage} and \emph{distribution}. 
For instance, given an event log $L=(E,\mi{Attr},\pi)$ and a set of time windows $W$, the coverage for any aspect $\mi{asp}$ that describes \emph{action} and any component $c$ can be defined as $\mi{cov}(\mi{asp},c)= \frac{1}{|E|}\sum_{w \in W} |f_{\mi{asp}}^{ev}(c,w)|$.
This metric indicates the proportion of events covered when analyzing aspect $\mi{asp}$ for component $c$.
High-level events referring to aspects and components that are more prevalent in the data and thus have higher coverage are generally more interesting.
Conversely, high-level events that arise from outlier values in the data can also be particularly meaningful. 
To effectively capture this, one could define an aggregated value from the list $\langle f_{\mi{asp}}^{val}(c,w_1), f_{\mi{asp}}^{val}(c,w_2), ..., f_{\mi{asp}}^{val}(c,w_{|W|})$ of aspect function values across different windows. 
This aggregated value helps to identify significant deviations or trends in the data, providing insights into rare but impactful high-level events.  

In Definition \ref{def:hle}, thresholds for detecting high-level events are uniquely set for each combination of aspect and component. 
This allows for a tailored approach to high-level event detection, ensuring that the thresholds are appropriately set for the specific characteristics of each aspect and component. One effective way to establish these thresholds is by setting them at the $p$th percentile of all values across the window set. 
This method guarantees that at least one high-level event will emerge for each aspect and component pair, capturing all significant deviations from typical values.
Having independent thresholds for different aspects is sensible because these aspects represent different process perspectives, making value comparisons between them often meaningless. However, using the same threshold for the same aspect across different components can be beneficial. 
This approach enables implicit comparisons of the same aspect values across various activities, resources, or segments, providing insights into the relative intensity or frequency of the aspect within different areas of the process.
Furthermore, the threshold function can be adapted so that it allows for the analysis of both unusually high and low occurrences.
By flipping the inequality sign in Definition \ref{def:hle}, one can shift the focus from detecting high levels of congestion or activity (typically seen as problematic) to identifying low levels. 
This analysis of low congestion or activity could be useful in scenarios where underutilization is a concern, helping to highlight areas of the process that may benefit from efficiency improvements or reallocation of resources. 

To enhance runtime efficiency, preprocessing steps like filtering the event log might be necessary, but these methods risk producing invalid results when detecting high-level events. For instance, projecting the event log onto specific activities or segments can simplify the data but may alter the original log by discarding certain activities and creating new activity pairs seen as segments in high-level events.
Moreover, removing entire traces from the event log can significantly impact the observed process load across all windows where the removed cases were active. 
This type of filtering could lead to a misrepresentation of process dynamics and result in inaccurate high-level events.

The intuition behind high-level events is that, similar to low-level events, they describe what happened and when it happened. While each high-level event is unique, detecting high-level events is interesting when there are recurring patterns. Analogous to the activity attribute of low-level events, one can determine a label that describes what happened when the high-level events emerged. 
In existing work, this label is a combination of both the aspect and the component related to the high-level event.
In the remainder, we assume that the aspect set $\mi{ASP}\subseteq\mathcal{U}_{\mi{asp}}$ analyzed for any given event log only contains aspects from the ones presented in this section.
\begin{definition}[High-level activity]\label{def:hla}
    Given event log $L=(E, \mi{Attr}, \pi)$, window set $W$, aspect set $\mi{ASP} \subseteq \mathcal{U}_{\mi{asp}}$ and threshold function $f_{\mi{thresh}}$, let $\mathcal{H}_{L,W,\mi{ASP},f_{\mi{thresh}}} \subseteq \mi{ASP} \times \mi{comp}(L) \times W$ be the set of high-level events detected w.r.t. $L$, $W$, $\mi{ASP}$ and $f_{\mi{thresh}}$.
    For any $h=(\mi{asp},c,w) \in \mathcal{H}_{L,W,\mi{ASP},f_{\mi{thresh}}}$, the corresponding \emph{high-level activity} is $\mi{act'}(h)=(\mi{asp},c) \in \mi{ASP} \times \mi{comp}(L)$.
    I.e., the high-level activity discards the time of emergence of the high-level event and refers (only) to both its aspect and its component.
\end{definition}
Note that discarding the time aspect is crucial for obtaining recurrent high-level behavior.
However, it is a design choice to consider both the aspect and the component as the high-level activity as opposed to choosing only the aspect or only the component.
\section{High-Level Event Connection: Propagation and Cascades}\label{sec:cascade and thread}
\subsection{Approaches to Computing Proximity Functions}
All events underlying each observed high-level event occur along the process run of their corresponding cases.
Hence, the high-level events that emerge throughout the time windows are implicitly caused by the progression of cases within the process. 
As tasks are handled sequentially for each case, whenever an event occurs, it may contribute to process overloading at the current and/or subsequent activity, resource, and/or segment.
The high-level events observed at a given point in time may reflect the propagation of effects along the process components of the cases that triggered the high-level events at earlier stages. 
Therefore, the high-level events observed throughout the process are interdependent. 
Establishing criteria to determine when two high-level events could be linked allows for the analysis of more complex high-level behavior, specifically sequences of high-level events and potentially recurring sequences of high-level activities. However, there is no single correct method of defining these criteria, and furthermore, it directly influences the interpretation of the resulting sequences of high-level events. Regardless of how it is computed, in existing work, the strength of connection is always a function that assigns a proximity value to any pair of high-level events.
\begin{definition}[Proximity function]\label{def:proximity}
Let $\mathcal{H}=\mathcal{H}_{L,W,\mi{ASP}, f_\mi{thresh}}$ be the set of high-level events obtained from event log $L$ with window set $W$, and aspects $\mi{ASP}$ with threshold function $f_\mi{thresh}$. 
Any function $\bowtie \in \mathcal{H}\times \mathcal{H} \rightarrow [0,1]$ is called a \emph{proximity function}, where for any two high-level events $\mi{h}_1, \mi{h}_2 \in \mathcal{H}$, ${\bowtie_L}(\mi{h}_1,\mi{h}_2)$ yields the \emph{proximity} between $\mi{h}_1$ and $\mi{h}_2$ with 0 being the farthest and 1 being the closest.
\end{definition}
In the following, we explore various options for proximity functions given a high-level event set $\mathcal{H}$ emerging from event log $L$, windows $W$, aspect set $\mi{ASP}$ and thresholds from $f_{\mi{thresh}}$.

The underlying component of each high-level event identifies the specific part of the process the high-level event relates to.
From this perspective, one can argue that certain component pairs are ``closer'' to each other than others;  for instance, a pair of activities that are always executed sequentially are close, while any resource is distant from activities it never executes. 
In \cite{hlem}, the proximity value is determined solely based on the underlying components of the high-level event pair.
A function $\mi{link} \in \mi{comp}(L) \times \mi{comp}(L) \rightarrow [0,1]$ is computed which assigns a value between 0 and 1 to each component pair.
Specifically, for any component $c \in \mi{comp}(L): \ \mi{link}(c,c)=1$ (each component is closest to itself), and for any two components $c_1, c_2 \in \mi{comp}(L): \ \mi{link}(c_1,c_2)=\mi{link}(c_2,c_1)$ (the order is irrelevant).
The detailed computation for all combinations of pairs of activities, resources, and segments can be found in \cite{hlem}.
The idea behind the \emph{link} value is that it increases with the number of events in the data that involve both components simultaneously.
Then, for any two high-level events $h_1=(\mi{asp_1},c_1,w_1), h_2=(\mi{asp_2},c_2,w_2) \in \mathcal{H}$, their proximity value $\bowtie(h_1,h_2)=\mi{link}(c_1,c_2)$ if there is no $w' \in W$ such that $w_1 < w' < w_2$ and $\bowtie(h_1,h_2)=0$ otherwise.
In other words, the proximity value of high-level events emerging in subsequent windows equals the \emph{link} value of their underlying components, and it is 0 if the high-level events occur farther apart from each other. 
It is important to note, however, that the \emph{link} value is determined by aggregating the entire event data.
Therefore, it is possible for two high-level events in subsequent windows to refer to components that are generally close in the process, but share few or no common events from those causing these specific high-level events.
One could also employ a stricter version of the \emph{link} function, assigning value 1 only to identical components, and 0 otherwise.
Consequently, pairs of high-level events would only receive a positive proximity value if they occur in subsequent windows and refer to the same activity, resource, or segment.

\begin{figure}[h]
\minipage{0.5\textwidth}
  \includegraphics[width=\linewidth]{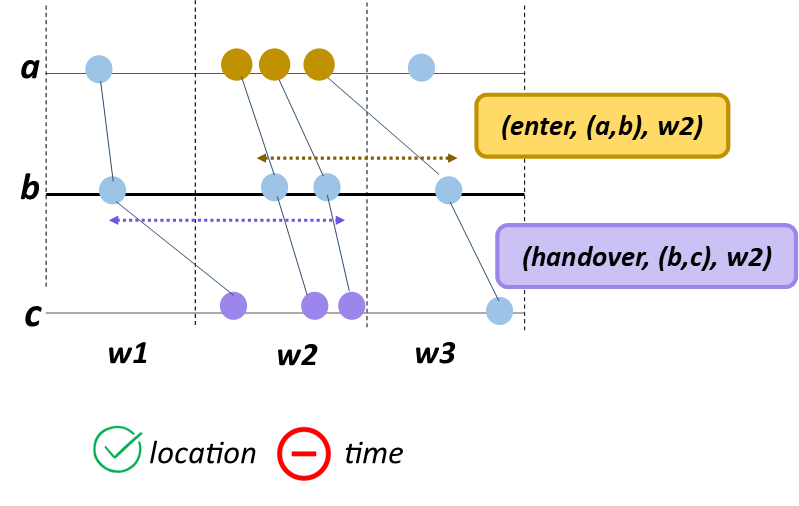}
\endminipage\hfill
\minipage{0.48\textwidth}%
  \includegraphics[width=\linewidth]{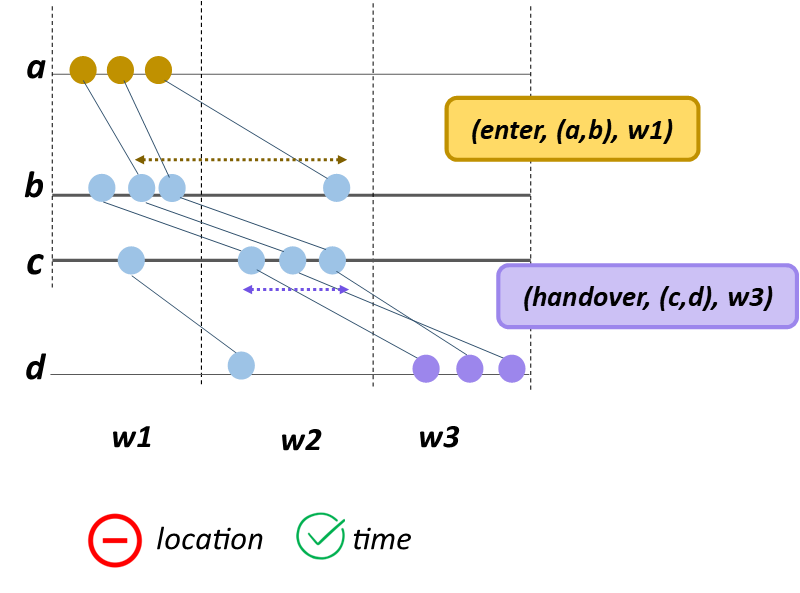}
\endminipage
 \caption{The high-level events $(\mi{enter},(a,b),w_2)$ and $(\mi{handover},(b,c),w_2)$ (left) satisfy location overlap (their corresponding segments share activity $b$ in the middle), but they do not satisfy time overlap.
 The high-level events $(\mi{enter},(a,b),w_1)$ and $(\mi{handover},(c,d),w_3)$ (left) satisfy time overlap (the time period in which the first high-level event ends encompasses the time period in which the second high-level event starts), but they do not satisfy location overlap (their corresponding segments are disconnected).}
 \label{fig: overlap time loc}
\end{figure}

The proximity value used between high-level events related to segment-based aspects is refined in \cite{interplay} to incorporate the set of underlying events involved.
Specifically, the proximity value can only be positive if the high-level events occur ``close'' in \emph{time} (referred to as \emph{time overlap}) and \emph{location} (referred to as \emph{location overlap}).
For instance, let $h_1=(\mi{asp_1},s_1,w_1), h_2=(\mi{asp_2},s_2,w_2) \in \mathcal{H}$ be two high-level events where $w_1 \leq w_2$, $s_1=(a,b), s_2=(c,d) \in S(L)$ are segments from the log and $\mi{asp_1}, \mi{asp_2}$ are segment-based aspects.
$h_1$ and $h_2$ satisfy \emph{location overlap} if and only if $b=c$.
Moreover, let $E_b(h_1) \subseteq E$ be the events that execute $b$ which either cause high-level event $h_1$ or are their immediate predecessors or successors w.r.t. segment $(a,b)$.
Let $\overrightarrow{w_b}(h_1)=[\mi{min}\{\mi{time}(e) \mid e \in E_b(h_1)\}, \mi{max}\{\mi{time}(e) \mid e \in E_b(h_1)\}]$ be the time period between the first and last event from $E_b(h_1)$.
Similarly, let $E_c(h_2) \subseteq E$ be the events that execute $c$ which either cause high-level event $h_2$ or are their immediate predecessors or successors w.r.t. segment $(c,d)$.
Let $\overrightarrow{w_c}(h_2)=[\mi{min}\{\mi{time}(e) \mid e \in E_c(h_2)\}, \mi{max}\{\mi{time}(e) \mid e \in E_c(h_2)\}]$ be the time period between the first and last event from $E_c(h_2)$.
$h_1$ and $h_2$ satisfy \emph{time overlap} if and only if $\overrightarrow{w_b}(h_1) \subseteq \overrightarrow{w_c}(h_2)$ or $\overrightarrow{w_c}(h_2) \subseteq \overrightarrow{w_b}(h_1)$.
In other words, $h_1$ and $h_2$ are close in space when $h_1$ ends where $h_2$ begins, and they are close in time when $h_1$ ends when $h_2$ begins.
Figure \ref{fig: overlap time loc} depicts two scenarios in which a pair of high-level events satisfy either time or location overlap, but not both.
In the left illustration of this figure, high-level events $h_1=(\mi{enter},(a,b),w_2)$ and $h=(\mi{handover},(b,c),w_2)$ share activity $b$ in the middle, and hence have location overlap with each other.
However, neither of the time periods in which the $h_1$ ends and $h_2$ begins is encompassed by the other, hence the time overlap criteria is not satisfied.
Conversely, in the example to the right, high-level events $h_1=(\mi{enter},(a,b),w_1)$ and $h=(\mi{handover},(b,c),w_3)$ satisfy time overlap, as the time period in which $h_2$ begins is encompassed by the time period in which $h_1$ ends.
However, these high-level events appear in disconnected segments, hence they do not satisfy location overlap.

\begin{figure}[h]
\centerline{\includegraphics[scale=0.5]{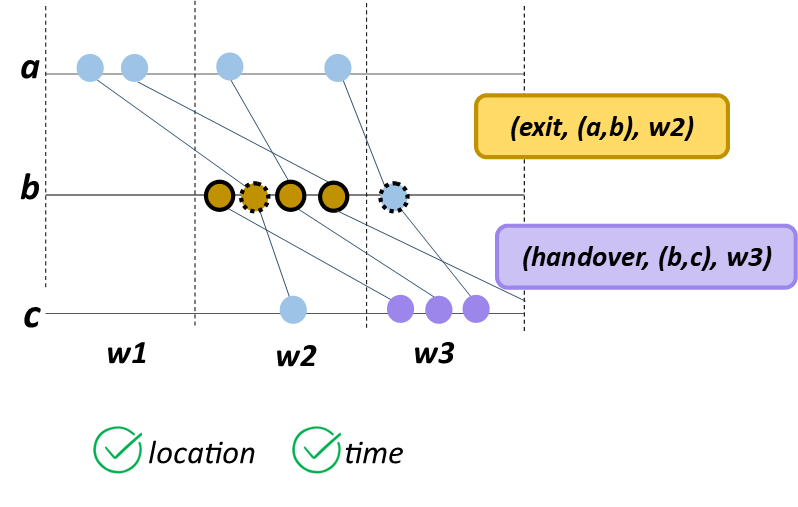}}
\caption{There are four cases behind the event set of high-level event $(\mi{exit},(a,b),w_2)$ and there are three cases behind the event set of the high-level event $(\mi{handover},(b,c),w_3)$. In total, there are five distinct cases behind these two high-level events, and three cases are part of both. Hence, the case overlap equals 3/5.}
\label{fig: overlap case}
\end{figure}

Additionally, the proximity value between high-level events that occur close in time and location depends on whether the underlying event sets belong to the same process instances.
Specifically, for any high-level event $h=(\mi{asp},c,w)$, its corresponding case set is $\mi{cases}(h)=\{\mi{case}(e) \mid e \in f_{\mi{asp}}^{ev}(c,w)\}$.
Hence, the \emph{case overlap} between the high-level events $h_1$ and $h_2$ is equal to $\mi{co}(h_1,h_2)=|\mi{cases}(h_1) \cap \mi{cases}(h_2)| / |\mi{cases}(h_1) \cup \mi{cases}(h_2)|$.
For example, for the two high-level events $h_1=(\mi{exit},(a,b),w_2)$ and $h=(\mi{handover},(b,c),w_3)$ from Figure \ref{fig: overlap case}, the case overlap is equal to 3/5.

Finally, $\bowtie(h_1,h_2) = \mi{co}(h_1,h_2)$ whenever $h_1$ and $h_2$ satisfy time and location overlap, and $\bowtie(h_1,h_2) = 0$ otherwise.
Note that by this definition, high-level events need not occur in subsequent time windows in order to have a positive proximity value.
Intuitively, a pair $(h_1,h_2)$ of high-level events related to segment-based aspects has a high proximity value whenever a high number of cases that ``participate'' in the first high-level event, also participate in the second high-level event immediately in the next step.


To extend the method from \cite{interplay} to cover high-level events related to all aspects at the activity, resource, or segment level, one approach is to incorporate the concept of \emph{instance overlap}. 
This type of proximity function, detailed below, relies on the underlying event sets responsible for triggering the high-level events, revealing the portion of events associated with the second high-level event that directly follow an event involved in the first high-level event.
\begin{definition}[Instance overlap]\label{def:instance}
Let $\mathcal{H}_{L,W,\mi{ASP}, f_\mi{thresh}}$ be the set of high-level events obtained from event log $L$ with window set $W$, and aspects $\mi{ASP}$ with threshold function $f_\mi{thresh}$. 
For any two high-level events $h_1=(\mi{asp_1},c_1,w_1), h_2=(\mi{asp_2},s_2,w_2) \in \mathcal{H}_{L,W,\mi{ASP}, f_\mi{thresh}}$, $\mi{io}(h_1,h_2) \in [0,1]$ yields the \emph{instance overlap} between $\mi{h}_1$ and $\mi{h}_2$, where 
\begin{align*}
    \mi{io}(h_1,h_2)= 
    \frac{|\{\mi{next}(e) \mid e \in f_{\mi{asp_1}}^{\mi{ev}}(c_1,w_1)\} \cap \{e \in f_{\mi{asp_2}}^{\mi{ev}}(c_2,w_2)\}| }
    {|\{\mi{next}(e) \mid e \in f_{\mi{asp_1}}^{\mi{ev}}(c_1,w_1)\} \cup \{e \in f_{\mi{asp_2}}^{\mi{ev}}(c_2,w_2)\}| }.
\end{align*}
\end{definition}
In other words, the instance overlap between a pair of high-level events quantifies the proportion of events in the second high-level event set that immediately follow events in the first high-level event set. 
Note that for segment-based aspects, the instance overlap integrates time, location, and case overlap simultaneously.
\subsection{Emergence and Interpretation of Cascades and Threads}
\begin{figure}[h]
\minipage{0.49\textwidth}
  \includegraphics[width=\linewidth]{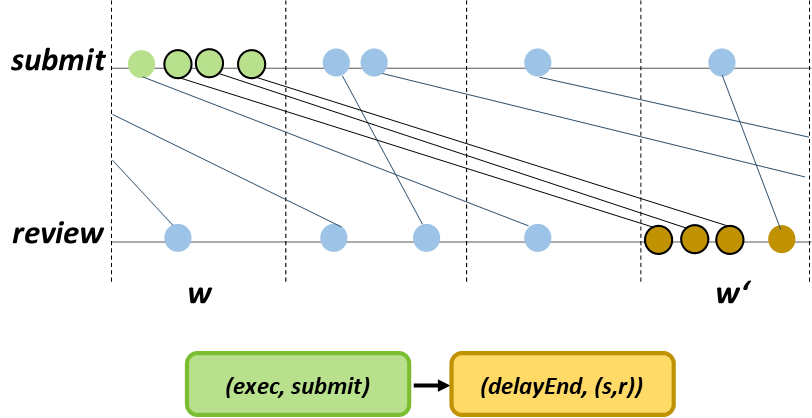}
\endminipage\hfill
\minipage{0.49\textwidth}%
  \includegraphics[width=\linewidth]{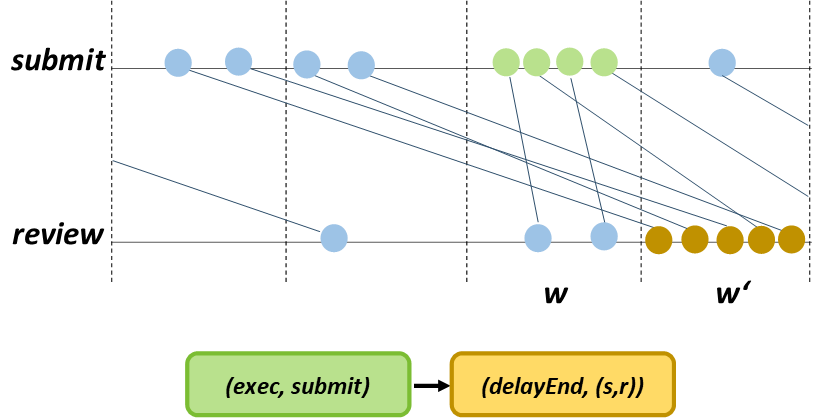}
\endminipage
 \caption{The illustration shows two different instantiations of the thread variant $\langle (\mi{exec},\mi{submit}),(\mi{delayEnd},(\mi{submit},\mi{review})) \rangle$.
 In the left example, high-level events $(\mi{exec},\mi{submit},w)$ and $(\mi{delayEnd},(\mi{submit},\mi{review}),w')$ occur in distant time windows. However, many events that directly follow the events in the first high-level event are also involved in the second.
 These high-level events are quite close regarding \emph{instance overlap}.
 In the right example, the same high-level events occur at subsequent time windows. 
 Moreover, the process components underlying them are very close in the process (\emph{review} always follows \emph{submit}).
 Hence, w.r.t. the \emph{link} method, they are very close to each other.
 Depending on the method used to define propagation, the resulting thread variant has a different interpretation.}
 \label{fig: proximity comparison}
\end{figure}

The proximity value between two high-level events addresses the direct relationship between them.
Using the proximity value, one can trace pairs of high-level events that are very close and thus obtain sequences of high-level events that reflect the emergence and dissolution of high-level behavior through time.
\begin{definition}[Propagation]\label{def:propagation}
    Let $\mathcal{H}=\mathcal{H}_{L,W,\mi{ASP}, f_\mi{thresh}}$ be the set of high-level events obtained from event log $L$ with window set $W$, and aspects $\mi{ASP}$ with threshold function $f_\mi{thresh}$. 
    Let $\bowtie$ be a proximity function and let $0 \leq \lambda \leq 1$.
    For any two high-level events $\mi{h}_1, \mi{h}_2 \in \mathcal{H}$, we say $h_1$ propagates to $h_2$ w.r.t. proximity function $\bowtie$ and proximity threshold $\lambda$ (denoted $h_1 \leadsto_{\bowtie, \lambda} h_2$) if and only if $\bowtie(h_1,h_2) \geq \lambda$.
    In particular, $\leadsto$ and $\lambda$ define a binary relation $\leadsto_{\bowtie, \lambda} \in \mathcal{H} \times \mathcal{H}$ over high-level event set $\mathcal{H}$ where $(h_1,h_2) \in \leadsto_{\bowtie,\lambda}$ if and only if $h_1 \leadsto_{\bowtie, \lambda} h_2$.
\end{definition}
In other words, a high-level event propagates to another high-level event whenever the pair is close enough, given a particular proximity function and a threshold for the proximity values.
\begin{definition}[Cascade]\label{def:cascade}
    Let $\mathcal{H}=\mathcal{H}_{L,W,\mi{ASP}, f_\mi{thresh}}$ be the set of high-level events obtained from event log $L$ with window set $W$, and aspects $\mi{ASP}$ with threshold function $f_\mi{thresh}$. 
    Let $\leadsto\in \mathcal{H} \times \mathcal{H}$ be a binary relation over the high-level event set.
    Let $\mi{casc}_{\leadsto} \in \mathcal{H} \rightarrow \mathbb{N}$ be a function that assigns a \emph{cascade identifier} to each high-level event w.r.t. $\leadsto$ such that for any two high-level events $h, h' \in \mathcal{H}$, the following holds:
    \begin{align*}
    \mi{casc}_{\leadsto}(h) = \mi{casc}_{\leadsto}(h') \ \Leftrightarrow \ 
    h \leadsto h' \ \vee \\
    \exists_{h_1,...,h_n \in \mathcal{H}} \ s.t. \ h \leadsto h_1 \ \wedge \ h_n \leadsto h' \ \wedge 
     \forall_{1 \leq i < n} \ h_i \leadsto h_{i+1}.
    \end{align*}
    For any cascade identifier $\mi{cid} \in \mi{rng}(\mi{casc}_{\leadsto})$, let $\mathcal{H}_{\mi{cid}}=\{h \in \mathcal{H} \mid \mi{casc}_{\leadsto}(h)=\mi{cid}\}$ be the high-level events that are assigned cascade id $\mi{cid}$.
    Moreover, let $W_{\mi{cid}} = \{w \in W \mid (\mi{asp},c,w) \in \mathcal{H}_{\mi{cid}}\}$ be the time windows in which those high-level events occur, and let $\langle w_1, ..., w_n\rangle \in W_{\mi{cid}}^*$ where $n=|W_{\mi{cid}}|$ and $\{w_i \mid 1\leq i \leq  n\}= W_{\mi{cid}}$ be the timely ordered sequence of those time windows.
    We define $\mathcal{C}(\mathcal{H}_{\mi{cid}})=\langle H_1,...,H_n \rangle \in (\mathcal{P}(\mathcal{H}_{\mi{cid}}))^*$ to be the \emph{cascade} related to cascade id $\mi{cid}$, where it holds that 
    {$\mathcal{H}_{\mi{cid}}=H_1 \cup ... \cup H_n$ and $\forall_{1 \leq i \leq n} \forall_{h \in H_i} \ h\in w_i$}. 
    $\mathfrak{C}_{\mathcal{H}, \leadsto} = \{\mathcal{C}(\mathcal{H}_{\mi{cid}}) \mid \mi{cid} \in \mi{rng}(\mi{casc}_{\leadsto})\} \subseteq (\mathcal{P}(\mathcal{H}))^*$ is the set of all cascades of high-level event set $\mathcal{H}$ w.r.t. $\leadsto$.
\end{definition}
Given a binary relation over high-level events determined by a proximity function and its corresponding threshold, any two high-level events are assigned the same cascade ID whenever there is a direct or indirect propagation from one high-level event to the other. 
When two high-level events have the same cascade identifier, we say they belong to the same cascade. 
A cascade is a sequence of high-level event sets ordered by the time window of their occurrence. 
This definition of cascades highlights that there may be multiple high-level events occurring in the same time window that propagate to (or from) a common high-level event in a future (or past) time window.
\begin{definition}[Thread]\label{def:thread}
    Let $\mathcal{H}=\mathcal{H}_{L,W,\mi{ASP}, f_\mi{thresh}}$ be the set of high-level events obtained from event log $L$ with window set $W$, and aspects $\mi{ASP}$ with threshold function $f_\mi{thresh}$. 
    Let $\leadsto \in \mathcal{H} \times \mathcal{H}$ be a binary relation over the high-level event set.
    Any sequence $\langle h_1,...,h_n \rangle \in \mathcal{H}^*$ such that for all $1 \leq i < n$ it holds that $h_i \leadsto h_{i+1}$ is called a \emph{thread} of $\mathcal{H}$ w.r.t. $\leadsto$. 
    Set $\mathfrak{T}_{\mathcal{H}, \leadsto} \subseteq \mathcal{H}^*$ contains all threads of $\mathcal{H}$ w.r.t. $\leadsto$.
\end{definition}
In other words, a thread is a sequence of high-level events connected to each other through propagation. 
By definition, all high-level events of the same thread belong to the same cascade. 
Moreover, one can further define \emph{maximal} threads (as opposed to \emph{subthreads}) which are threads of maximal length, as no other high-level event can be added to the start or the end of the corresponding propagation sequence.
{\begin{definition}[Cascade variant, Thread variant]\label{def:variant}
    Let $\mathcal{H}=\mathcal{H}_{L,W,\mi{ASP}, f_\mi{thresh}}$ be the set of high-level events obtained from event log $L$ with window set $W$, and aspects $\mi{ASP}$ with threshold function $f_\mi{thresh}$. 
    Let $\leadsto \in \mathcal{H} \times \mathcal{H}$ be a binary relation over the high-level event set and let $\mathfrak{C}_{\mathcal{H}, \leadsto}$ and $\mathfrak{T}_{\mathcal{H}, \leadsto}$ be the sets of all cascades and threads of $\mathcal{H}$ and $\leadsto$ respectively.
    For any cascade $\mathcal{C} = \langle H_1,...,H_n \rangle \in \mathfrak{C}_{\mathcal{H}, \leadsto} $, we define its \emph{variant} to be the sequence $\mathcal{C}' = \langle H_1',...,H_n'\rangle$ where for all $1 \leq i \leq n$: $H_i'=\{\mi{act}'(h) \mid h \in H_i\}$.
    I.e., a cascade variant is the sequence of its high-level event sets projected onto their high-level activity sets.
    Similarly, for any thread $\mathcal{T} = \langle h_1,...,h_m \rangle \in \mathfrak{T}_{\mathcal{H}, \leadsto}$, we define its \emph{variant} to be the sequence $\mathcal{T}' = \langle \mi{act}'(h_1),...,\mi{act}'(h_m)\rangle$ as the sequence of its high-level events projected onto their high-level activities.
    Sets $\mathfrak{C}'_{\mathcal{H}, \leadsto}$ and $\mathfrak{T}'_{\mathcal{H}, \leadsto}$ are the sets of all cascade variants and thread variants of $\mathcal{H}$ and $\leadsto$ respectively.
\end{definition}}
Figure \ref{fig: proximity comparison} presents two different examples of how the pairs of high-level events $(\mi{exec},\mi{submit},w)$ and $(\mi{delayEnd},(\mi{submit},\mi{review}),w')$ can occur. 
In the left example, the second high-level event takes place three windows later than the first, with three of the four events involved in the second high-level directly following the four events involved in the first high-level event. 
Employing \emph{instance overlap} with any threshold $\leq 3/5$ consequently connects these two events. 
The corresponding thread variant is $\langle (\mi{exec},\mi{submit}),(\mi{delayEnd},(\mi{submit},\mi{review})) \rangle$. 
The right example illustrates how the same thread variant could manifest when using \emph{link}—the second high-level event occurs in the window immediately following the first. 
While the segment $(\mi{submit},\mi{review})$ is close to activity $\mi{submit}$ (note that in the citizenship process from Figure \ref{fig: process}, \emph{review} always follows \emph{submit}), there is minimal overlap among the event sets involved in the high-level events. 
This example underscores the varying interpretations of emerging thread variants depending on the defined propagation method.

\begin{figure}[h]
\minipage{0.45\textwidth}
  \includegraphics[width=\linewidth]{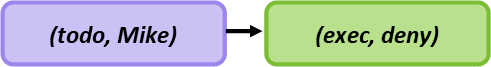}
\endminipage\hfill
\minipage{0.45\textwidth}%
  \includegraphics[width=\linewidth]{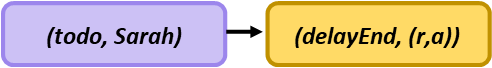}
\endminipage
 \caption{Two possible thread variants in the citizenship application process: the \emph{todo} aspect, indicating increased workload for Mike and Sarah; the \emph{delayEnd} aspect, highlighting delays in case processing between \emph{r: review} and \emph{a:accept}; and the \emph{exec} aspect, identifying unusually frequent activity execution. Using instance overlap, these thread variants suggest higher denial rates for cases managed by Mike during busy periods and extended approval times for cases assigned to Sarah when she is overwhelmed.}
 \label{fig: io thread 1 3}
\end{figure}

Figure \ref{fig: io thread 1 3} illustrates two thread variants related to the hidden behaviors in the citizenship application process, as discussed in Section \ref{sec:motivation}. 
The aspect \emph{todo} helps identify when Mike and Sarah are suddenly overwhelmed with more work than usual, while \emph{delayEnd} aids in detecting when cases exiting a segment within a brief period have encountered delays. 
Additionally, \emph{exec} is useful for detecting when an activity is executed more frequently than usual within a time window. 
For example, using instance overlap, one could determine that cases assigned to Mike in large quantities are more likely to be denied, or that cases assigned to Sarah in large quantities tend to experience longer waits until final approval.

\begin{figure}[h]
\centerline{\includegraphics[scale=0.33]{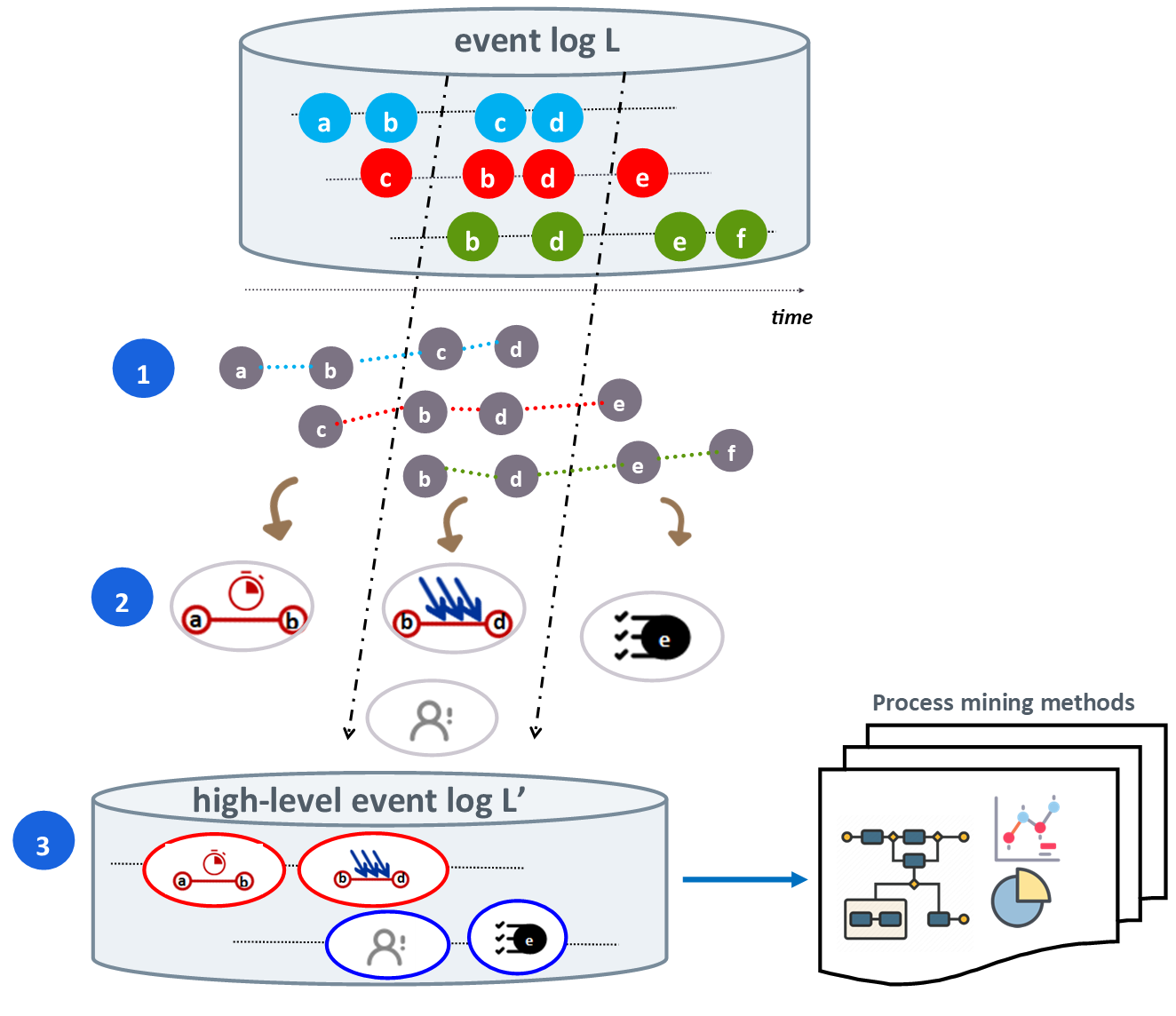}}
\caption{A visualization of the approach from \cite{hlem}:
On top, the input event data showing the runs of three process instances (blue, red and green). The time scope is split into time windows (step 1), and the events within the same time window may produce certain process patterns. These patterns are captured as high-level events (step 2) which are then collected into a high-level event log (step 3).}
\label{fig: framework}
\end{figure}

Note that in \cite{interplay}, each thread is referred to as an \emph{episode}, whereas each thread variant is referred to as the \emph{high-level path}.
In \cite{hlem}, a high-level event log is generated where the events correspond to the high-level events.
A visualization of the framework is shown in Figure \ref{fig: framework}.
The activity attribute corresponds to the high-level activity, while the time attribute is obtained from the window where the high-level event emerged (can be set to the start or the end of the time window).
High-level events are organized into high-level cases which correspond to the cascades emerging using a proximity measure and a chosen threshold for the high-level event pairs.
One can exploit the event log format of the high-level event log to apply existing process mining tools and techniques for further insights.
However, it is important to note that events of the same high-level case in the new log are not necessarily totally ordered.
High-level events occurring in the same time window and belonging to the same cascade obtain identical timestamps.
As most process mining techniques assume that events of a case can be totally ordered through time, one can introduce a new total order on the high-level event set, such as using the lexicographical order of the activity labels.
Even when applied to the high-level log with the lexicographical total order, the results of some existing process mining techniques can still be insightful.
\section{Interplay with the Underlying Instances}\label{sec:overlap}
As outlined in the definition of high-level events, a specific set of events is always involved whenever such behavior emerges. 
Moreover, each event belongs to a process instance, which, in turn, has its own set of attributes. 
Case-level attributes can be defined as event attributes that hold the same value for all events belonging to the same case. 
On one hand, the characteristics of a case may exacerbate the emergence of high-level behavior; for example, a demanding case can tie up resources for longer periods. 
On the other hand, the outcome of a case can also be influenced by high-level behavior occurring throughout its execution; for instance, a case may not receive the necessary attention if it happens to enter the process during a busy period. 
There is undoubtedly an interplay between the high-level behavior that arises in the process and the cases that trigger it. 
The method introduced in \cite{interplay} explores this interplay by detecting which patterns of high-level behavior emerge surprisingly often from specific case types. 
Possessing this knowledge offers numerous advantages. 
Depending on the case property at hand, one can adjust the process for specific types of cases to avoid expected but undesirable high-level problems, or make better online predictions about the progress of a case based on its involvement in specific patterns of high-level behavior.

Next, we provide a brief overview of the method introduced in \cite{interplay}.
As discussed in the previous section, all emerging high-level events are associated with segment-based aspects and are connected to each other using a proximity function that considers factors such as time overlap, location overlap and case overlap.
Specifically, a high-level event $h_1$ that occurs on segment $(a_1,b_1)$ propagates to a later high-level event $h_2$ on segment $(a_2,b_2)$ under the following conditions: 
1) $b_1=a_2$ (location overlap), 
2) the time period of the $b_1$-related events for $h_1$ either encompasses or is encompassed by the time period of the $a_2$-related events for $h_2$ (time overlap), and 
3) there is a significant overlap in the cases involved in both high-level events, e.g., $\geq 0.5$ (case overlap).
For each individual high-level event $h$, given the set of the events that caused its occurrence, one can determine their corresponding set of cases.
These are the cases that ``participate'' in $h$.
As high-level events are connected into longer sequences through propagation, the concept of participating cases can be extended to sequences. 
The following definition covers the concept of participating cases for all types of high-level events, threads, and thread variants.
To compute the propagation relation, one could use \emph{instance overlap} as defined in Definition \ref{def:instance} for the proximity function $\bowtie$, combined with any threshold $\lambda \in [0,1]$ .
For high-level events related to segment-based aspects and thread variants created using time, location and case overlap, the definition below aligns with the one used in \cite{interplay}.
\begin{definition}[Participating cases]\label{def:participation}
    Let $\mathcal{H}=\mathcal{H}_{L,W,\mi{ASP}, f_\mi{thresh}}$ be the set of high-level events obtained from event log $L$ with window set $W$, and aspects $\mi{ASP}$ with threshold function $f_\mi{thresh}$. 
    Let $\leadsto \in \mathcal{H} \times \mathcal{H}$ be a binary relation over the high-level event set and let $\mathfrak{T}_{\mathcal{H}, \leadsto}$ and $\mathfrak{T}'_{\mathcal{H}, \leadsto}$ be the sets of all threads and thread variants of $\mathcal{H}$ and $\leadsto$ respectively.
    For each high-level event $h =(\mi{asp},c,w) \in \mathcal{H}$, the \emph{participating cases} of $h$ are $\mi{cases}(h)=\{\mi{case}(e) \mid e \in f_{\mi{asp}}^{\mi{ev}}(c,w)\}$.
    Moreover, for any thread $\mathcal{T} \in \mathfrak{T}_{\mathcal{H}, \leadsto} $, the \emph{participating cases} of $\mathcal{T}$ are $\mi{cases}(\mathcal{T})=\bigcap_{h \in \mathcal{T}} \mi{cases}(h)$.
    Finally, for any thread variant $\mathcal{T}' \in \mathfrak{T}'_{\mathcal{H}, \leadsto} $, the \emph{participating cases} of $\mathcal{T}'$ are $\mi{cases}(\mathcal{T}')=\bigcup_{\substack{\mathcal{X} \in \mathfrak{T}_{\mathcal{H}, \leadsto} \\ s.t. \ \mathcal{X}' = \mathcal{T}'}} \mi{cases}(\mathcal{X})$.
\end{definition}
To sum up, the participating cases of an individual high-level event consist of the cases corresponding to the event set involved in that high-level event. 
Within each thread, the participating cases encompass those that are involved in all high-level events contained in the thread. 
Similarly, for each thread variant, the participating cases constitute the union of all cases involved in threads whose corresponding variants equal the given variant.

\begin{figure}[h]
\centerline{\includegraphics[scale=0.5]{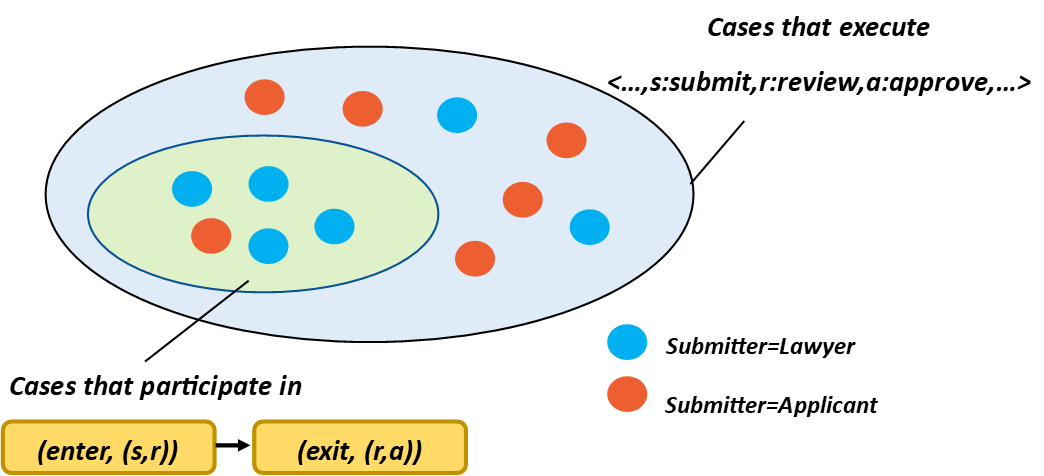}}
\caption{An illustration from \cite{interplay} demonstrates the method: each dot represents an application, with orange dots signifying those submitted by the applicants themselves and blue dots representing those submitted by lawyers. 
The large blue circle encompasses all cases where the activity sequence $\langle \mi{submit}, \mi{review}, \mi{approve} \rangle$ appears in their trace. 
Within this, the green circle highlights a subset of cases that, throughout their processing, participate in the thread variant $\langle (\mi{enter},(\mi{submit},\mi{review})),(\mi{exit},(\mi{review},\mi{approve})) \rangle$. 
This diagram reveals that applications submitted by lawyers are more prevalent among the cases involved in this specific thread variant.}
\label{fig: interplay}
\end{figure}

The interplay between case types and high-level behavior (thread variants) can be assessed using the correlation value obtained from $\chi^2$.
This statistical test measures the disparity between observed and expected frequencies for each combination of values of two categorical variables. 
In our scenario, one can measure a correlation value for each pair of thread variant and case attribute (see Table 1).
The first categorical variable corresponds to the case attribute itself. 
To incorporate numerical case attributes, such as throughput time, binning methods can be applied. 
This case attribute partitions the set of cases into categories based on the possible values of the attribute. 
The second categorical attribute reflects case participation in the given thread variant, dividing the cases into two groups: ``participating'' and ``non-participating.''
The null hypothesis states that there is no relationship between case participation in a given thread variant and the chosen case attribute. 
A correlation is considered statistically significant, leading to the rejection of the null hypothesis, if the corresponding $p$-value is less than $0.05$. 
It is important to clarify the definition of ``non-participating'' cases for each thread (variant). 
These cases form the control group against which the correlation is evaluated. 
Notably, not all cases not participating are part of this control group. 
Specifically, cases not involved in any activities, resources, or segments related to the high-level events of the thread are, by definition, not part of the participating group.
It is more sensible to compare the participating cases with a control group of cases which could have been participating w.r.t. the process components the high-level events concern.
In \cite{interplay}, any emerging sequence of high-level activities (thread variant) $\langle h_{(a_1,b_1)},...,h_{(a_n,b_n)} \rangle$ concerns a sequence of underlying segments in the process.
Due to the location overlap criteria used for propagation, it holds that $b_i=a_{i+1}$ for all $1 \leq i < n$.
Hence, a corresponding sequence of activities $\langle a_1,b_1,b_2,...,b_{n-1},b_n \rangle$ can always be obtained as the control-flow sequence underlying the thread variant.
The non-participating cases are defined as the set of cases which execute that activity sequence throughout their traces, but never participate in the high-level behavior at hand (the thread variant).
Hence, the control group are the cases which are not participating, but could have participated from a control-flow perspective (see Figure \ref{fig: interplay}).
\begin{table}[h]\label{table:chi}
\caption{Given some thread variant $\mathcal{T'}$, the participating cases and non-participating cases are further split based on the chosen categorical attribute values (here: category 1 and category 2). The correlation between the attribute and the thread variant is computed using the $\chi^2$ test of independence on the row partition (the chosen case-level attribute) and on the column partition ((non-)participation in the thread variant).}
\centering
\begin{tabular}{| c | c | c | c |}  
\hline
\textbf{Case-level attribute} & \textbf{Participating cases of $\mathcal{T}'$} & \textbf{Non-participating cases of $\mathcal{T}'$}\\ \hline
category 1          & $n_1$                        & $n_2$                                      \\ \hline
category 2          & $n_3$                        & $n_4$                                      \\ \hline
                    & $n_1+n_3=|\mi{cases}(\mathcal{T}')|$              & $n_2+n_4$                 \\ \hline
\end{tabular}
\end{table}

\begin{figure}[h]\label{fig:path1}
\centerline{\includegraphics[scale=0.7]{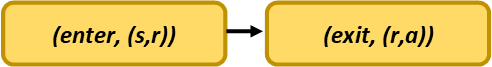}}
\caption{An illustration of a thread variant that shows a significant correlation with the submitter of citizenship applications: applications which tend to be submitted and approved in batches are mostly submitted by lawyers. This indicates that lawyers might strategically choose optimal submission times and that these applications often receive faster, grouped approvals due to prior checking.
}
\label{fig: path1}
\end{figure}

Assume that in the citizenship process described in Section \ref{sec:motivation}, each application can be submitted either by the applicant themselves or by a lawyer representing them. 
Suppose that the thread variant from Figure \ref{fig: path1} shows a notable correlation with the submitter of the application, with those submitted by lawyers being disproportionately represented among the participating cases. 
This suggests that applications submitted by lawyers tend to be entered into the process in large numbers over brief time intervals and are often approved in batches. 
This pattern may indicate that lawyers are adept at identifying strategical times for submission.
Furthermore, these applications frequently receive simultaneous approval, likely because the prerequisites for citizenship have already been thoroughly checked by the lawyers.

\begin{figure}[h]\label{fig:path2}
\centerline{\includegraphics[scale=0.7]{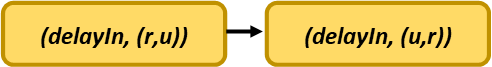}}
\caption{An illustration of a thread variant which is more commonly associated with applications submitted directly by the applicants themselves, indicating that these individuals often face delays in updating their documents and during the review stage.}
\label{fig: path2}
\end{figure}

Furthermore, consider that the thread variant depicted in Figure \ref{fig: path2} is predominantly associated with applications submitted directly by the applicants themselves. 
This observation suggests that among applicants required to update their documents and subsequently wait for a review, those who submit their own applications are more likely to experience delays in both of these process stages. 

To adapt the method for a broader scenario, one could define the underlying component sequence for thread variants referring not only to segments but also to activities and resources. 
Instance overlap from Definition \ref{def:instance} reflects the fraction of events shared ``in the middle'' of any two high-level events. 
These shared events could be used to establish the underlying component sequence and eventually the control group for any thread variant.

It is important to emphasize that the number of threads can grow exponentially with the number of cascades, making it infeasible to compute case partitions for every thread variant. 
One approach, as employed in \cite{interplay}, is to focus on threads where the fraction of shared cases from the first to the last high-level event is high, similar to the threshold used for case overlap.

An extension to the method in \cite{interplay} would be to consider not only thread variants, but also cascade variants.
This would need defining the participating cases of cascades.
Requiring that a case must be involved in all high-level events of a cascade to be considered as participating may be overly strict; depending on the high-level event types that emerge in the same window within the same cascade, it can be impossible for any case to participate in all of them.

Even when restricting the analysis to a subset of thread/cascade variants, there could still be numerous variants showing significant correlation with a given case property.
Ideally, these variants would be ranked from most interesting to least interesting.
The ranking could be based solely on the variants set and be independent of the chosen case property.
For instance, the ranking of cascade/thread variants could be determined based on factors such as \emph{size}, \emph{frequency}, and \emph{reach}.
The size of a variant is determined by its number of high-level activities. 
Larger variants are less likely to arise randomly, making them more ``distinguishable'' and thus contributing to a higher rank.
Frequency, on the other hand, reflects how often a variant occurs. 
More frequent variants are less likely to represent random behavior, thereby enhancing their rank.
The \emph{reach} of a variant is defined by the ratio between participating and non-participating cases. 
A higher proportion of participating cases compared to non-participating ones indicates a more prevalent behavior within the underlying process components, leading to a higher rank.
Note that size, frequency and reach capture distinct aspects of variant \emph{interestingness}.
A weighted average of these factors could yield a single rank value for each variant.
With a predefined ranking of variants, users could opt to view only the $n$ highest-ranked variants among those exhibiting significant correlation with a given case property.
\section{Exploring Process Robustness}\label{sec:robustness}
An interesting future research direction involves leveraging high-level event mining to analyze process \emph{robustness} and \emph{resilience}.
Robustness refers to the ability of a system to maintain its expected performance or functionality despite variations in input, workload or conditions.
It relates to the system's capacity to handle typical fluctuations, variations in workload, or minor disruptions without experiencing significant adverse effects.
Resilience is the capacity of a system to absorb disruptions and recover quickly by returning to a stable state or optimal performance.
In other words, \emph{robustness} relates to the system's ability to remain functional in the face of disruption, whereas \emph{resilience} relates to the system's ability to bounce back to a normal state after some disruption.
In the context of process mining and system-level analysis, disruptions can manifest in the form of process load fluctuations. For instance, adverse effects of process load spikes can include increased \emph{delays} for process instances and heightened \emph{workload} for resources. 
Analyzing process robustness and resilience requires identifying disruptions and measuring their adverse effects.

Next, we will briefly outline a method that treats disruptions as sudden load spikes at the activity level and evaluates their impact on case delay. 
In process mining, it is typically assumed that the process owner has no control over incoming case frequency and is subject to its variability. 
As the process unfolds, certain activities must be performed by specific resources in a particular order, and cases proceed through the process stages accordingly. 
At various time intervals, sudden spikes in incoming cases may occur. 
In these situations, cases may accumulate at initial activities, but spikes can also manifest elsewhere in the process.
When multiple cases arrive at the same process stage simultaneously, they strain resource capacity, posing a threat to process stability. 
If one would visualize the size of the queue at a given activity throughout different time windows, the spikes in the graph would reveal the situations where the queue abruptly expands due to simultaneous case entries.

Consider the process from event log $L=(E,\mi{Attr},\pi)$ across time windows $W$. 
A case enters the queue at activity $a$ upon completing the preceding activity and remains in the queue until $a$ is executed. If numerous cases arrive at activity $a$ simultaneously, the queue can suddenly become too large. 
We define a \emph{disruption} at activity $a \in A(L)$ during $w \in W$ whenever the queue length at $a$ during $w$ is large, with the cases enqueued during $w$ contributing significantly to it.
\begin{definition}[Disruption]\label{def:disruption}
    Let $L=(E,\mi{Attr},\pi)$ be an event log $L$ and let $W$ be a set of time windows.
    For any activity $a \in A(L)$, let $q_a \in \mathbb{N}$ be a threshold for the queue length at $a$ and let $p_a \in [0,1]$ be a threshold for the portion of the recently enqueued cases in the queue.
    Given set $D \subseteq A(L) \times W$ of all disruptions, we say there is a \emph{disruption at $a$ during $w$} (denoted $(a,w) \in D $) whenever $f_{\mi{queue}}^{\mi{val}}(a,w) \geq q_a$ (the queue length is large) and $f_{\mi{enqueue}}^{\mi{val}}(a,w) / f_{\mi{queue}}^{\mi{val}}(a,w) \geq p_a$ (the newly enqueued cases make up a big portion of the queue).
\end{definition}

The effects of the disruption may manifest in various ways in the process. 
For instance, cases entering the same queue at the same time or shortly after the disruption may experience prolonged waiting times. 
Additionally, in response to the sudden process load at the activity, resources may need to work more than usual.

For any activity $a$ and window $w$, cases that are in the queue during $w$ spend an average time equal to $\mi{wt}(a,w)=\frac{1}{|\mi{f_{\mi{queue}}^{\mi{ev}}(a,w)}|} \sum_{e \in f_{\mi{queue}}^{\mi{ev}}(a,w)} \big(\mi{time}(\mi{next}(e)) - \mi{time}(e)\big)$ waiting until $a$. 
Note that the impact of a disruption onto case waiting time may not be immediate.
To analyze this impact, one could monitor the waiting time for several time windows following $w$. 
To identify which time windows may still be affected by the disruption, for any disruption $(a,w) \in D$, one can determine a set $W_{(a,w)} \subseteq W$ of corresponding ``potentially affected'' time windows, referred to as the \emph{resolution scope}.
\begin{definition}[Resolution scope]\label{def:resolution}
    Let $L=(E,\mi{Attr},\pi)$ be an event log $L$, let $W$ be a set of time windows, and let $D \subseteq A(L) \times W$ be the set of disruptions.
    For any $w \in W$, let $w-$ denote its preceding window.
    For any activity $a \in A(L)$ and window $w \in W$, $\mi{takeover}(a,w)=|\mi{f_{\mi{queue}}^{\mi{ev}}(a,w)} \setminus \mi{f_{\mi{enqueue}}^{\mi{ev}}(a,w)}|$ is the \emph{takeover} at $a$ during $w$, and it is the number of cases in the queue at $a$ during $w$ that have entered the queue previous to $w$.
    Similarly, $\mi{takeoverRatio}(a,w)=\mi{takeover}(a,w) / |\mi{f_{\mi{queue}}^{\mi{ev}}(a,w)}|$ is the \emph{takeover ratio} at $a$ during $w$, and it is the proportion of cases in the queue at $a$ during $w$ that have entered the queue previous to $w$.
    For any $a \in A(L)$, let $t_a \in \mathbb{N}$ and $tr_a \in [0,1]$ be thresholds related to the takeover and the takeover ratio at $a$.
    For any disruption $(a,w) \in D$, set $W_{(a,w)} \subseteq W$ is its \emph{resolution scope}, where for any $w' \in W$:
    \begin{align*}
        w'\in W_{(a,w)} \Leftrightarrow w'=w \ \vee \ \\
        w'- \in W_{(a,w)} \ \wedge \  \mi{takeover}(a,w')\geq t_a \ \wedge \ \mi{takeoverRatio}(a,w') \geq tr_a.
    \end{align*}
\end{definition}
In other words, the resolution scope of a disruption pertains to the windows following the window of the disruption. 
Any window within the resolution scope indicates that the process during that time window may still be experiencing the effects of the disruption.
Intuitively, the process is impacted by the disruption at the time window it occurs, and it remains affected in the subsequent time window if the queue at that new time window predominantly consists of a high number of cases from the previously affected time window.

Using the resolution scope, one can assess how much the average waiting time of cases active during the affected time windows differs from that of cases that remain unaffected by the disruption. 
In the following definition, we propose a method to measure waiting time robustness for any given activity.
\begin{definition}[Waiting Time Robustness]\label{def:wt robustness}
    Let $L=(E,\mi{Attr},\pi)$ be an event log $L$, let $W$ be a set of time windows, and let $D \subseteq A(L) \times W$ be the set of disruptions.
    For any activity $a \in A(L)$, let $\mi{events}(D,a)=\bigcup_{(a,w) \in D} \bigcup_{w \in W_{(a,w)}} \linebreak {\{e \in  f_{\mi{queue}}^{\mi{ev}}(a,w)\}}$ be the set of events indicating cases queueing at $a$ during time windows that are affected by disruptions.
    Moreover, let $\mi{wt}(D,a) = \frac{1}{|\mi{events}(D,a)|} \linebreak \sum_{e \in \mi{events}(D,a)} {\big(\mi{time} (\mi{next}(e)) - \mi{time}(e)\big)}$ be their average waiting time until $a$.
    Conversely, let $\overline{\mi{events}}(D,a)= \{e \in E \mid \mi{next}(e)=a\} \setminus \mi{events}(D,a)$ be the set of events indicating cases queueing at $a$ during time windows that are unaffected by disruptions, and let $\overline{\mi{wt}}(D,a) = \frac{1}{|\overline{\mi{events}}(D,a)|} \sum_{e \in \overline{\mi{events}}(D,a)} \big(\mi{time}(\mi{next}(e)) - \mi{time}(e)\big)$ be their average waiting time until $a$.
    We define $R_{\mi{wt}}(D,a)= \frac{\mi{wt}(D,a)}{\overline{\mi{wt}}(D,a)}$ to be the \emph{waiting time robustness indicator at activity $a$} as the waiting time ratio between cases that are affected by disruptions and the cases that are unaffected by disruptions.
\end{definition}
Note that $R_{\mi{wt}}(D,a)$ indicates the extent to which cases experience prolonged waiting times when they are involved in a disruption at activity $a$.
A value close to 1 suggests that the impact of these disruptions on waiting times is minimal. 
Any value greater than 1 signifies a negative effect on waiting times due to these disruptions.
It is worth noting that values smaller than 1 might occur if resources handle a significantly higher number of $a$ activities than usual whenever sudden large queues form at $a$.
\section{Conclusion and Directions for Future Work}\label{sec:conclusion}
In this report, we elaborated on the concept of high-level event mining. 
This is an important topic because many performance and compliance problems can only be understood by zooming out and considering patterns involving multiple events and cases. 
We formalized the notion of high-level events and provided examples of high-level behavior concerning congestion related aspects arising at the activity, resource and segment level. 
We also showed how to connect the different high-level events leading to the notion of cascades and threads. 
All high-level artifacts can be related to the initial events and cases causing their emergence.
Through high-level event mining, one can explore recurring cascades and threads, analyze their correlation with the types of the underlying cases and resources, and moreover, address process robustness and resilience.

Other promising paths for future research involve providing online predictions and support for active cases. 
In the context of high-level events, this could mean forecasting potential emergent high-level events in the upcoming time window. 
Such predictions could be based on patterns of high-level behavior observed previously in the process. 
It is essential that the event data is divided into training and test datasets to learn these behavior patterns, with the split aligning with a temporal point that separates historical cases (training data) from future cases (test data). 
Otherwise, removing cases from the time period used for training affects the resulting high-level behavior that is detected.

Another challenging aspect for future work is determining a suitable framing function which splits the time scope into time windows. 
Currently, the time window set is typically determined after users select a suitable window size based on domain expertise. 
However, when analyzing various activities, resources, and segments, no single window size may adequately capture high-level behavior across all components. 
One potential future direction could involve adapting existing methodologies to accommodate variable window sizes for different components. 
Alternatively, discarding the temporal partition altogether and instead utilizing methods that detect high-level events based on event accumulation within a coordinate space (e.g., time and component) could be explored.

Lastly, an interactive tool could greatly assist users in exploring the outcomes of existing methodologies. 
This tool could allow users to select the window size, relevant aspects, components, case properties, and various thresholds, facilitating a more comprehensive exploration of the results.

%
%
%
%
%
\bibliographystyle{splncs04}
\bibliography{report.bib}
%
%




\end{document}